\pdfoutput=1
\documentclass[a4paper,11pt]{article}

\usepackage{lineno}
\usepackage{jheppub}
\usepackage[utf8]{inputenc}
\usepackage{amsmath}
\allowdisplaybreaks

\usepackage{amsthm}
\usepackage{amssymb}
\usepackage{enumitem}   
\usepackage[english]{babel}
\usepackage{url}
\usepackage{mathtools}
\usepackage{bbold}
\usepackage{slashed}
\usepackage{multirow}
\usepackage{lipsum}
\usepackage{xcolor}
\usepackage{float}
\usepackage{revsymb}
\usepackage{tcolorbox}

\usepackage{nicematrix}

\newcommand{\bm}[1]{\boldsymbol{#1}}

\begin{document}


\preprint{MS-TP-26-07}
\title{Nonlinear physics of axion inflation}

\author[a,b]{Oleksandr~Sobol,}
\author[a]{Richard~von~Eckardstein,}
\author[a]{Elias~Koch,}
\author[a]{Svetlana~Gurevich,}
\author[a]{Uwe Thiele,}
\author[a,c]{and Kai~Schmitz}

\affiliation[a]{Institute for Theoretical Physics, University of M\"unster,\\
Wilhelm-Klemm-Stra{\ss}e 9, 48149 M\"{u}nster, Germany}

\affiliation[b]{Physics Faculty, Taras Shevchenko National University of Kyiv,\\
64/13, Volodymyrska Street, 01601 Kyiv, Ukraine}

\affiliation[c]{Kavli IPMU (WPI), UTIAS, The University of Tokyo,\\
5-1-5 Kashiwanoha, Kashiwa, Chiba 277-8583, Japan}

\emailAdd{richard.voneckardstein@uni-muenster.de}
\emailAdd{gurevics@uni-muenster.de}
\emailAdd{e\_koch12@uni-muenster.de}
\emailAdd{kai.schmitz@uni-muenster.de}
\emailAdd{oleksandr.sobol@uni-muenster.de}
\emailAdd{u.thiele@uni-muenster.de}


\abstract{An axion-like field coupled to an Abelian gauge field provides one of the simplest inflationary models that is free from the eta problem and possesses an efficient reheating mechanism. For sufficiently large coupling, this system enters a regime of strong gauge-field backreaction, exhibiting rich and intricate dynamics. In this work, we employ a semi-analytical method, the gradient-expansion formalism, to perform a comprehensive parameter scan and determine the precise conditions under which backreaction sets in.
Previous studies have shown that the Anber--Sorbo solution, in which the potential-gradient force acting on the axion is balanced by Hubble friction and gauge-field backreaction, is unstable. Here, we broaden the parameter space and identify a new region in which the Anber--Sorbo solution remains stable despite strong backreaction. Although our analysis is restricted to a homogeneous axion field and to perturbations that depend only on time, we expect that this stability property can be extrapolated to generic time- and space-dependent perturbations. This newly identified region therefore represents a distinct type of backreaction\,---\,\textit{stable backreaction}\,---\,which may not be accompanied by the rapid growth of perturbations.
We further investigate the nonlinear behavior of solutions in the backreaction regime in a toy model (de Sitter, constant potential slope, no axion gradients), identifying a supercritical Hopf bifurcation at the onset of instability, a nontrivial limit cycle in the unstable regime, and burst-like oscillatory dynamics. Finally, we present a more stringent criterion for the onset of (unstable) backreaction, based on crossing the instability threshold, and apply this criterion to two benchmark inflationary models.}



\maketitle


\section{Introduction}
\label{sec:introduction}

The inflationary paradigm has proven highly successful in addressing the ``fine-tuning'' puzzles of the Hot Big Bang theory, such as the horizon, flatness, and monopole problems~\cite{Starobinsky:1980te,Guth:1980zm,Linde:1981mu,Starobinsky:1982ee,Albrecht:1982wi,Linde:1983gd}. In addition, it naturally generates primordial scalar~\cite{Mukhanov:1981xt,Mukhanov:1982nu,Guth:1982ec,Hawking:1982cz,Bardeen:1983qw} and tensor~\cite{Grishchuk:1974ny,Starobinsky:1979ty,Rubakov:1982df,Fabbri:1983us,Abbott:1984fp} perturbations, which become imprinted in the cosmic microwave background (CMB) anisotropies and in the large-scale structure (LSS) of the Universe, and also give rise to a stochastic gravitational-wave background. Observations of the CMB~\cite{Planck:2018jri,BICEP:2021xfz,Tristram:2021tvh,AtacamaCosmologyTelescope:2025nti,SPT-3G:2025bzu} and LSS~\cite{Reid:2009xm,Cabass:2022wjy} place stringent constraints on the parameters of inflationary models, requiring the inflaton field $\phi$ to evolve in an extremely flat potential $V(\phi)$; for reviews, see Refs.~\cite{Martin:2013tda,Chluba:2015bqa,Martin:2018ycu,Durrer:2020fza,Martin:2024qnn}. Achieving such flatness, however, is challenging in a complete quantum theory of inflation, since radiative corrections can easily spoil the flatness of the tree-level potential. This difficulty is known in the literature as the eta problem, where eta denotes the second slow-roll (SR) parameter, $\eta_V=M_{\mathrm{P}}^2 (V''/V)$, which becomes large in the presence of these corrections.%
\footnote{In this work, we adopt natural units by setting $\hbar=c=1$, so that the reduced Planck mass is given by $M_{\mathrm{P}}=2.43\times 10^{18}\,\text{GeV}$.}
Therefore, any viable inflationary model must include a mechanism that suppresses these corrections, at least during the epoch when the perturbation modes relevant for observations cross the Hubble horizon. One possible strategy is to minimize the couplings between the inflaton and other fields. However, this approach is not ideal, since the inflaton must eventually decay into Standard Model particles to reheat the Universe after inflation. A more robust alternative is to protect the flatness of the inflaton potential through an underlying symmetry that prevents large radiative corrections.

This idea was implemented in the natural inflation model proposed in Refs.~\cite{Freese:1990rb,Adams:1992bn}. In this framework, the shift symmetry of a pseudoscalar (axion-like) inflaton $\phi$ protects its potential from large radiative corrections. Although the original model is now ruled out by CMB observations~\cite{Planck:2018jri,BICEP:2021xfz,Tristram:2021tvh}, several nonminimal extensions~\cite{Kim:2004rp,DAmico:2017cda,Pajer:2013fsa} remain viable~\cite{Peloso:2015dsa,Copeland:2022lkp,Greco:2024ngr}. Moreover, this class of models naturally accommodates a shift-symmetric coupling between the inflaton and gauge fields of the form $\propto \phi F^{\mu\nu}\tilde{F}_{\mu\nu}$, where $F^{\mu\nu}$ is the gauge-field strength tensor and $\tilde{F}_{\mu\nu}$ its dual. Both Abelian and non-Abelian gauge fields can couple in this way,%
\footnote{While in this work we consider only the coupling to an Abelian gauge field, there exists a substantial body of literature exploring axion--gauge-field dynamics in the non-Abelian case; see, e.g., Refs.~\cite{Adshead:2012kp,Maleknejad:2018nxz,Domcke:2018rvv,Domcke:2019lxq,Dimastrogiovanni:2025snj,Ishiwata:2025wmo}. Another class of models assumes that the axion-like field does not serve as the inflaton but instead acts as a spectator field; see Refs.~\cite{Barnaby:2012xt,Namba:2015gja,Peloso:2016gqs,Caravano:2024xsb} for models featuring an axion spectator field coupling to an Abelian gauge field and Refs.~\cite{Dimastrogiovanni:2016fuu,Ishiwata:2021yne,Iarygina:2023mtj} for models featuring an axion spectator field coupling to non-Abelian gauge fields.}
allowing the inflaton to interact efficiently with Standard Model fields without introducing large radiative corrections to its potential.

The axion--vector coupling leads to efficient gauge-field production through the tachyonic enhancement of vacuum gauge-field modes of a single circular polarization, resulting in field configurations with nonzero helicity~\cite{Anber:2006xt,Anber:2009ua}. These fields may be regarded as precursors of large-scale magnetic fields in the Universe~\cite{Turner:1987bw,Garretson:1992vt,Anber:2006xt,Durrer:2010mq,Caprini:2014mja,Bamba:2014vda,Fujita:2015iga,Adshead:2016iae,Fujita:2019pmi,Shtanov:2019civ,Shtanov:2019gpx,Sobol:2019xls} or as dark photons~\cite{Bastero-Gil:2021wsf,Domcke:2021yuz,Lysenko:2025sdo}, which are viable dark-matter candidates. They may also lead to the generation of the baryon asymmetry via the chiral anomaly~\cite{Anber:2015yca,Adshead:2016iae,Jimenez:2017cdr,Domcke:2019mnd,Domcke:2022kfs,Brandenburg:2023aco}. During inflation these fields can produce charged particles through the Schwinger effect~\cite{Domcke:2018eki,Domcke:2019qmm,Sobol:2019xls,Gorbar:2021rlt,Gorbar:2021zlr,Cado:2022pxk,Fujita:2022fwc,Gorbar:2021ajq,Bastero-Gil:2023mxm,Bastero-Gil:2023htv,vonEckardstein:2024tix,Iarygina:2025ncl}.
The presence of gauge fields induces large and non-Gaussian scalar perturbations~\cite{Barnaby:2010vf,Barnaby:2011vw,Barnaby:2011qe,Sorbo:2011rz,Ferreira:2014zia,Ferreira:2015omg,Peloso:2016gqs,Domcke:2020zez,Durrer:2024ibi,Galanti:2024jhw,Talebian:2025jeg}, with the possible formation of primordial black holes during the post-inflationary radiation-dominated epoch~\cite{Linde:2012bt,Bugaev:2013fya,Cheng:2015oqa,Garcia-Bellido:2016dkw,Garcia-Bellido:2017aan,Domcke:2020zez,Ozsoy:2023ryl}. In addition, tensor perturbations are generated, giving rise to a stochastic gravitational-wave background~\cite{Cook:2011hg,Domcke:2016bkh,Garcia-Bellido:2016dkw,Garcia-Bellido:2017aan,Garcia-Bellido:2023ser,Corba:2024tfz,vonEckardstein:2025oic,vonEckardstein:2025elq,Barbon:2025wjl,Teuscher:2025jhq,Corba:2025reo}. Finally, the coupling of an axion-like inflaton to gauge fields provides an efficient reheating mechanism after inflation~\cite{Adshead:2015pva,Figueroa:2017qmv,Adshead:2018doq,Cuissa:2018oiw,Adshead:2019igv,Adshead:2019lbr}. This extremely rich phenomenology of Abelian axion inflation calls for its thorough investigation.

For sufficiently small couplings, gauge-field production can be described perturbatively by tracking the time evolution of individual gauge-field Fourier modes in momentum space. By contrast, for larger coupling values, the gauge field begins to affect the background inflationary dynamics, and the system enters the strong-backreaction regime, in which all modes are coupled nonlinearly. Originally, it was hoped that the additional friction induced by the gauge-field backreaction could balance the axion potential gradient and sustain slow roll even for a steep potential~\cite{Anber:2009ua}. However, many subsequent studies~\cite{Notari:2016npn,Cheng:2015oqa,DallAgata:2019yrr,Domcke:2016bkh,Gorbar:2021rlt} demonstrated that an exact balance between the potential gradient force and the gauge-field friction is not possible. In fact, the gauge-field response is always delayed with respect to changes in the inflaton velocity~\cite{Notari:2016npn,Domcke:2020zez}. This delay induces rapid oscillations of both the inflaton velocity and the gauge-field energy density, preventing the system from remaining in a SR regime. Moreover, Refs.~\cite{Peloso:2022ovc,vonEckardstein:2023gwk} showed that even if the system is prepared in a state where the inflaton potential gradient is exactly compensated by the gauge-field backreaction term (the so-called Anber--Sorbo state), any small perturbation quickly destroys the initial force balance and triggers fast oscillations. Consequently, the Anber--Sorbo solution was argued to be unstable throughout the strong-backreaction regime.

A somewhat related observation is that, in the strong-backreaction regime, scalar perturbations induced by the presence of gauge fields become much larger than the usual vacuum fluctuations. This was first demonstrated in semianalytical studies~\cite{Barnaby:2010vf,Barnaby:2011qe,Barnaby:2011vw} employing perturbative approaches; see also Refs.~\cite{Durrer:2024ibi,Galanti:2024jhw,Barbon:2025wjl} for recent investigations of this topic. Moreover, several lattice simulations of axion inflation show rapid growth of axion inhomogeneities (i.e., axion gradients) in the regime of strong gauge-field backreaction~\cite{Caravano:2022epk,Figueroa:2023oxc,Sharma:2024nfu,Figueroa:2024rkr,Jamieson:2025ngu,Lizarraga:2025aiw}.
An interesting question is therefore whether the occurrence of fast oscillations of the axion velocity in the homogeneous description directly corresponds to the rapid growth of axion gradients in the fully inhomogeneous picture. A definitive answer requires a comprehensive scan of parameter space using lattice techniques. However, even with current technology, such a scan is computationally very expensive. Therefore, preliminary results from semianalytical approaches based on a homogeneous approximation can provide valuable insights by addressing the following questions: (i)~where exactly is the boundary in parameter space that delineates the strong-backreaction regime; (ii)~what is the rate of growth of perturbations (depending only on time, not on spatial coordinates) within the backreaction regime; and (iii)~is there a region in which the perturbations (at least in a homogeneous description) do not grow. In this paper, we carry out such an analysis using the gradient-expansion formalism (GEF) and provide answers to the questions listed above.

The remainder of this work is organized as follows. In Sec.~\ref{sec:axion-inflation}, we review the basic features of the model of axion inflation under consideration. Section~\ref{sec:const_back} is devoted to a simplified setup of constant-roll axion inflation in pure de Sitter space. We first discuss the stationary solution corresponding to a constant instability variable~$\xi$ in Sec.~\ref{subsec:stationary-sol}. In Sec.~\ref{subsec:Lyapunov}, we then analyze its stability with respect to small time-dependent perturbations using the linearized gradient-expansion formalism. Based on the instability of the stationary solution, we introduce a new criterion for the strong-backreaction regime, which is more stringent than existing criteria in the literature. In Sec.~\ref{subsec:criterion}, we discuss this new criterion in greater detail and derive analytical fit functions for the threshold values of the relevant parameters at the onset of the backreaction regime. Next, in Sec.~\ref{subsec:bifurcation}, we examine how the system responds to a gradual increase of the axion--vector coupling for a fixed potential shape, and show that the variable~$\xi$ undergoes a supercritical Hopf bifurcation at the boundary of the instability region. The properties of the system deep inside the instability region are studied in Sec.~\ref{subsec:bursting}. Finally, in Sec.~\ref{sec:realistic}, we use the GEF to study the dynamics of gauge-field production in a two realistic inflationary models and illustrate how our new criterion for the strong backreaction operates there.
Section~\ref{sec:conclusion} contains our conclusions. In Appendix~\ref{app:GEF}, we review the main ingredients of the gradient-expansion formalism. Appendix~\ref{app:comp-previous} compares the parameter region explored in this work to those considered in previous studies.

\newpage

\section{Axion inflation}
\label{sec:axion-inflation}


In the model of axion inflation, a pseudoscalar inflaton field $\phi$ interacts with a $U(1)$ gauge field $A_\mu$ through a Chern--Simons-type coupling. The total action describing this system is given by
\begin{equation}
\label{action-1}
S[\phi,\,A_\mu]=\int d^4x\,\sqrt{-g}\left[-\frac{1}{2}g^{\mu\nu}\partial_\mu\phi\partial_\nu\phi - V(\phi) - \frac{1}{4}F_{\mu\nu}F^{\mu\nu} - \frac{\beta\phi}{4M_\mathrm{P}}F_{\mu\nu}\tilde{F}^{\mu\nu}\right],
\end{equation}
where $g_{\mu\nu}$ is the spacetime metric, $g=\operatorname{det}g_{\mu\nu}$ is its determinant, $V(\phi)$ denotes the potential of the inflaton field, $F_{\mu\nu}=\partial_\mu A_\nu - \partial_\nu A_\mu$ represents the field strength tensor, and $\tilde{F}^{\mu\nu}=\varepsilon^{\mu\nu\alpha\beta}F_{\alpha\beta}/(2\sqrt{-g})$ is its dual, with $\varepsilon^{\mu\nu\alpha\beta}$ the totally antisymmetric Levi-Civita symbol normalized as $\varepsilon^{0123}=1$. The parameter $\beta$ quantifies the dimensionless coupling between the axion and the gauge field.  

By varying the action \eqref{action-1}, we derive the equations of motion for both $\phi$ and $A_\mu$:
\begin{align}
\label{KGF}
\frac{1}{\sqrt{-g}}\partial_\mu\left(\sqrt{-g}\,g^{\mu\nu}\partial_\nu\phi\right) - \frac{dV}{d\phi} - \frac{\beta}{4M_\mathrm{P}}F_{\mu\nu}\tilde{F}^{\mu\nu} &= 0\,, \\
\label{Maxwell-cov}
\frac{1}{\sqrt{-g}}\partial_\mu\left(\sqrt{-g}\,F^{\mu\nu}\right) + \frac{\beta}{M_\mathrm{P}}\tilde{F}^{\mu\nu}\partial_\mu\phi &= 0\,,
\end{align}
where we used that the dual field tensor satisfies the Bianchi identity:
\begin{equation}
\label{Maxwell-Bianchi}
\frac{1}{\sqrt{-g}}\partial_\mu\left(\sqrt{-g}\,\tilde{F}^{\mu\nu}\right)=0\,.
\end{equation}

The inflationary background is assumed to be homogeneous and isotropic, described by the spatially flat Friedmann--Lema\^{i}tre--Robertson--Walker (FLRW) metric. Written in terms of cosmic time $t$ or conformal time $\eta$, the line element takes the form
\begin{equation}
\label{eq:FLRW-metric}
ds^2=g_{\mu\nu}\,\mathrm{d}x^\mu\mathrm{d}x^\nu=-\mathrm{d}t^2+a^2(t)\,\mathrm{d}\bm{x}^2 = a^2(\eta)\left[-\mathrm{d}\eta^2+\mathrm{d}\bm{x}^2\right],
\end{equation}
where $a(t)$ denotes the scale factor; the number of $e$-foldings, $N\equiv\ln a$, is also often used as a timelike variable. Correspondingly, we also postulate that the axion field is homogeneous in space and solely dependent on time, i.e., $\phi = \phi(t)$.%
\footnote{This is a very good approximation during slow-roll inflation, as the typical amplitude of inhomogeneities is several orders of magnitude smaller compared to the homogeneous field. However, as demonstrated by semianalytical approaches~\cite{Peloso:2022ovc,vonEckardstein:2023gwk} and lattice simulations~\cite{Figueroa:2023oxc,Figueroa:2024rkr,Sharma:2024nfu}, in the case of strong gauge-field backreaction, inflaton inhomogeneities tend to proliferate, and within a few $e$-foldings, they may attain a magnitude sufficient to exert a substantial influence on the overall dynamics of the model. Having in mind these limitations of the homogeneous approximation, we still adopt it throughout this work and perform a comprehensive parameter scan of the axion inflation model. In the analysis of our results, we clearly state which of them remain robust against the impact of inhomogeneities and which may be strongly modified and need to be cross-checked by lattice simulations.}
In the temporal gauge, $A_\mu=(0,\bm{A})$, the physical electric and magnetic fields in the comoving frame are defined as
\begin{equation}
\label{eq:E-and-B-through-A}
\bm{E}=-\frac{1}{a}\frac{\partial\bm{A}}{\partial t},\qquad
\bm{B}=\frac{1}{a^2}\operatorname{rot}\bm{A}.
\end{equation}
The field strength tensor and its dual can then be expressed in terms of $\bm{E}$ and $\bm{B}$ as follows:
\begin{equation}
F_{0i}=-aE^i,\qquad F_{ij}=a^2\varepsilon_{ijk}B^k,\qquad
\tilde{F}_{0i}=-aB^i,\qquad
\tilde{F}_{ij}=-a^2\varepsilon_{ijk}E^k,
\end{equation}
where $\varepsilon_{ijk}$ denotes the Levi-Civita symbol in three dimensions.

The energy--momentum tensor derived from \eqref{action-1} takes the form
\begin{equation}
\label{T}
T^\mu_\nu=-\partial^\mu\phi\partial_\nu\phi-F^{\mu\lambda}F_{\nu\lambda}+\delta^\mu_\nu\left[\frac{1}{2}\partial_\alpha\phi\partial^\alpha\phi+V(\phi)+\frac{1}{4}F_{\alpha\beta}F^{\alpha\beta}\right],
\end{equation}
which, for a homogeneous inflaton field, yields the total energy density
\begin{equation}
\label{rho}
\rho=\langle T^0_0\rangle=\frac{1}{2}\dot{\phi}^2+V(\phi)+\frac{1}{2}\langle\bm{E}^2+\bm{B}^2\rangle,
\end{equation}
where an overdot denotes the derivative with respect to cosmic time $t$ and $\langle\cdots\rangle$ denotes the quantum expectation value during inflation. The Hubble rate is determined by the Friedmann equation:
\begin{equation}
\label{Friedmann}
H^2\equiv \Big(\frac{\dot{a}}{a}\Big)^2=\frac{\rho}{3M_\mathrm{P}^2}\,.
\end{equation}

Finally, the equations of motion \eqref{KGF}--\eqref{Maxwell-Bianchi} can be expressed in terms of three-vector quantities:
\begin{align}
\label{KGF-2}
\ddot{\phi}+3H\dot{\phi}+V'(\phi) &=\frac{\beta}{M_\mathrm{P}}\langle\bm{E}\cdot\bm{B}\rangle,\\
\label{Maxwell_1}
\dot{\bm{E}}+2H\bm{E}-\frac{1}{a}\operatorname{rot}\bm{B}+\frac{\beta}{M_\mathrm{P}}\dot{\phi}\bm{B} &=0,\\
\label{Maxwell_2}
\dot{\bm{B}}+2H\bm{B}+\frac{1}{a}\operatorname{rot}\bm{E}&=0,\\
\label{Maxwell_3}
\operatorname{div}\bm{E}=0,\qquad
\operatorname{div}\bm{B}&=0.
\end{align}
Together, Eqs.~\eqref{Friedmann}--\eqref{Maxwell_3} form a self-consistent system describing the coupled evolution of the homogeneous inflaton, the scale factor, and the gauge field during axion inflation. The only remaining missing ingredient is a prescription for computing the vacuum expectation values that enter the Friedmann and Klein--Gordon equations. To establish this, we must consider the quantized vector field.

Let us now switch to momentum space and write the gauge-field operator as
\begin{equation}
\label{quantized_A}
\hat{\bm{A}}(t,\bm{x})=\int\frac{d^{3}\bm{k}}{(2\pi)^{3/2}}\sum_{\lambda=\pm}\left[\bm{\epsilon}^{\lambda}(\bm{k})\hat{a}_{\bm{k},\lambda}A_{\lambda}(t,k)e^{i\bm{k}\cdot\bm{x}}+\bm{\epsilon}^{\lambda *}(\bm{k})\hat{a}_{\bm{k},\lambda}^{\dagger}A_{\lambda}^{*}(t,k)e^{-i\bm{k}\cdot\bm{x}} \right] \,,
\end{equation}
where $A_{\lambda}(t,k)$ is the  mode function, $\bm{\epsilon}^{\lambda}(\bm{k})$ is the polarization vector, and $\hat{a}_{\bm{k},\lambda}$ ($\hat{a}^{\dagger}_{\bm{k},\lambda}$) are the annihilation (creation) operators for gauge-field modes with momentum $\bm{k}$ and circular polarization $\lambda=\pm$, and $k=|\bm{k}|$. The polarization vectors have the following properties:
\begin{equation}
\bm{k}\cdot\bm{\epsilon}^{\lambda}(\bm{k})=0,\quad \bm{\epsilon}^{\lambda *}(\bm{k})=\bm{\epsilon}^{-\lambda}(\bm{k}), \quad [i\bm{k}\times\bm{\epsilon}^{\lambda}(\bm{k})]=\lambda k \bm{\epsilon}^{\lambda}(\bm{k}), \quad \bm{\epsilon}^{\lambda *}(\bm{k})\cdot\bm{\epsilon}^{\lambda'}(\bm{k})=\delta^{\lambda\lambda'} \,.
\end{equation}
The creation and annihilation operators obey the canonical commutation relations
\begin{equation}
[\hat{a}_{\bm{k},\lambda},\,\hat{a}^{\dagger}_{\bm{k}',\lambda'}]=\delta_{\lambda\lambda'}\delta^{(3)}(\bm{k}-\bm{k}')\,.
\end{equation}
Finally, the mode functions obey the Wronskian normalization condition
\begin{equation}
\label{eq:Wronskian-normalization}
    A_{\lambda}(t,k) \dot{A}^{\ast}_{\lambda}(t,k)-A^{\ast}_{\lambda}(t,k)\dot{A}_{\lambda}(t,k) = \frac{i}{a}\, ,
\end{equation}
which ensures the canonical Bohr--Heisenberg commutation relations between the field $\hat{\boldsymbol{A}}(t,\boldsymbol{x})$ and its conjugate momentum.

For the electric and magnetic fields expressed as in Eq.~\eqref{eq:E-and-B-through-A}, the Maxwell equations~\eqref{Maxwell_2}--\eqref{Maxwell_3} are identically satisfied, while Eq.~\eqref{Maxwell_1} leads to an equation of motion for the mode functions,
\begin{equation}
\label{eq:mode-equation-physical-time}
\ddot{A}_{\lambda}(t,k)+H\dot{A}_{\lambda}(t, k)+\left[\frac{k^{2}}{a^{2}}-2\lambda\frac{k}{a}H\xi\right] A_{\lambda}(t,k) =0 \,,
\end{equation}
where we introduced the quantity $\xi$ as usual:
\begin{equation}
\label{xi}
    \xi \equiv \frac{\beta \dot{\phi}}{2HM_{\mathrm{P}}} \,.
\end{equation}
It evolves slowly during slow-roll inflation, since the time variation of both the Hubble rate $H$ and the inflaton velocity $\dot{\phi}$ are suppressed by the slow-roll parameters. It is therefore instructive to assume, for the moment, that this quantity is constant. Since the two terms in square brackets in Eq.~\eqref{eq:mode-equation-physical-time} scale differently with the scale factor, one can distinguish two regimes in the evolution of the solutions of this equation. For any fixed $k$, at sufficiently early times when $k/a \gg 2H|\xi|$, the first term dominates over the second, and the equation of motion reduces to its form in the absence of interactions. Rewriting the equation in conformal time, $\eta$, which is particularly convenient for studying the evolution of conformally invariant fields such as a free massless Abelian gauge field, the equation takes an oscillator-like form:
\begin{equation}
    \label{A_2}
    \frac{\partial^2 A_\lambda(\eta, k)}{\partial\eta^2}+k^2 A_\lambda(\eta, k)\simeq0 \qquad\text{for}\qquad k/aH\gg 2|\xi|\, . 
\end{equation}
Its positive-frequency solution, normalized according to Eq.~\eqref{eq:Wronskian-normalization}, reads
\begin{equation}
\label{BD-condition}
    A_\lambda(\eta, k)\simeq\frac{1}{\sqrt{2k}}e^{-ik\eta} \qquad\text{for}\qquad k/aH\gg 2|\xi|\,.
\end{equation}
We will use this expression to impose the boundary condition for the gauge-field mode function at early times, or, in other words, \textit{deep inside the horizon},%
\footnote{For a free massless Abelian gauge field, Eq.~\eqref{A_2} implies that there is no relevant energy scale at which the evolution changes qualitatively: the field always oscillates in conformal time with a constant frequency. However, the presence of an axial coupling introduces such a scale, $2H|\xi|$, which defines a physically relevant threshold, or ``effective horizon''. If the physical momentum of the mode is much larger than this scale, i.e., if the reduced physical wavelength $\lambda_{\mathrm{phys}}/(2\pi)$ is much smaller than the effective horizon size $(2H|\xi|)^{-1}$, the mode oscillates as in the free case. In the opposite regime, $\lambda_{\mathrm{phys}}/(2\pi) \gg (2H|\xi|)^{-1}$, the evolution of the mode is dominated by the interaction.}
the so-called Bunch--Davies boundary condition~\cite{Bunch:1978yq}.

On the other hand, if $k/a \ll 2H|\xi|$, the second term in square brackets in Eq.~\eqref{eq:mode-equation-physical-time} dominates. In contrast to the term $k^2/a^2$, which is always positive and therefore only leads to oscillatory solutions, this term has opposite signs for modes with different circular polarizations. For $\lambda = \operatorname{sign}\xi$, the term is negative and gives rise to a tachyonic instability, resulting in a strong amplification of the corresponding mode function. Conversely, for the mode with $\lambda = -\operatorname{sign}\xi$, the term is positive and no tachyonic instability occurs. We thus conclude that superhorizon modes of one circular polarization undergo tachyonic enhancement when their momentum satisfies
\begin{equation}
\label{eq:tachyonic-condition}
    k<k_{\mathrm{thr}}(t) = 2a(t)H(t)|\xi(t)|
\end{equation}
at the moment of consideration. The instability is thus controlled by the variable $\xi$, which we refer to as \textit{the instability variable}.%
\footnote{In much of the literature, it is often referred to as the ``instability parameter,'' since, in the absence of gauge-field backreaction, it effectively acts as an external parameter controlling the intensity of gauge-field production. In the present work, however, this quantity is treated as one of the dynamical variables of the system. For this reason, we adopt the term ``instability variable'' throughout our paper.}%
\ This mechanism underlies the production of helical gauge fields during axion inflation.

Finally, we are now able to compute the vacuum expectation values of the gauge-field bilinear quantities Eqs.~\eqref{rho} and \eqref{KGF-2}. Using the definition of the electric and magnetic fields in Eq.~\eqref{eq:E-and-B-through-A} and substituting the explicit form of the quantized vector potential \eqref{quantized_A}, one can straightforwardly derive for the Klein--Gordon equation
\begin{equation}
    \label{eq:Klein-Gordon-final}
    \ddot{\phi}+3H\dot{\phi}+V'(\phi) =\frac{\beta}{M_\mathrm{P}}\langle\bm{E}\cdot\bm{B}\rangle = -\frac{\beta}{4\pi^2 a^3 M_\mathrm{P}}\int\limits_0^{k_{\mathrm{h}}(t)}\!\!\!dk \sum\limits_{\lambda=\pm}\lambda k^3\frac{\partial}{\partial t}\big|A_{\lambda}(t,k)\big|^2\, ,
\end{equation}
and for the Friedmann equation
\begin{align}
    H^2 &= \frac{1}{3M_{\mathrm{P}}^2}\Big[\frac{1}{2}\dot{\phi}^2 + V(\phi) +\frac{1}{2}\langle\bm{E}^2 + \bm{B}^2\rangle\Big]\nonumber\\
    &=\frac{1}{3M_{\mathrm{P}}^2}\bigg[\frac{1}{2}\dot{\phi}^2 + V(\phi) + \frac{1}{4\pi^2 a^4}\int\limits_0^{k_{\mathrm{h}}(t)}\!\!\! dk\,\sum\limits_{\lambda=\pm}\Big(k^4\big|A_{\lambda}\big|^2 +k^2 a^2\big|\dot{A}_{\lambda}\big|^2\Big)\bigg]\, .
    \label{eq:Friedmann-final}
\end{align}
Here, the momentum scale $k_{\mathrm{h}}(t)$ in Eqs.~\eqref{eq:Klein-Gordon-final} and \eqref{eq:Friedmann-final} serves as a finite upper integration limit. This cutoff is introduced to separate the physically relevant modes, which are enhanced due to the axial coupling to inflaton, from vacuum-like oscillatory modes. We define $k_{\mathrm{h}}(t)$ as the maximum, over all times $t' \leq t$, of the threshold momentum $k_{\mathrm{thr}}$ [see Eq.~\eqref{eq:tachyonic-condition}] that triggers the tachyonic instability in one of the polarization modes, namely
\begin{equation}
    k_{\mathrm{h}}(t) = \underset{t' \leq t}{\operatorname{max}}\: k_{\mathrm{thr}}(t') = \underset{t' \leq t}{\operatorname{max}} \:\big|2a(t')H(t')\xi(t')\big|\, .
\label{eq:cutoff-momentum}
\end{equation}

Therefore, Eqs.~\eqref{eq:Klein-Gordon-final} and \eqref{eq:Friedmann-final}, together with the mode equation~\eqref{eq:mode-equation-physical-time} supplemented by the boundary condition \eqref{BD-condition}, constitute the complete set of equations describing the dynamics of axion inflation with a homogeneous axion field in a mixed position--momentum representation. In particular, Eqs.~\eqref{eq:Klein-Gordon-final} and \eqref{eq:Friedmann-final} are integro--differential equations, with integrals taken over a continuous range of Fourier modes. To obtain their solution, one must discretize momentum space and simultaneously solve a large number of mode equations~\eqref{eq:mode-equation-physical-time} on a momentum lattice. At each time step, the contributions of all relevant modes are then summed in order to evolve the axion field and the scale factor through Eqs.~\eqref{eq:Klein-Gordon-final} and \eqref{eq:Friedmann-final}.
Throughout this work, we refer to this procedure as \textit{the mode-by-mode (MbM) approach}. It has been employed extensively in the literature to study the dynamics of axion inflation~\cite{Notari:2016npn,Cheng:2015oqa,DallAgata:2019yrr} as well as various phenomenological applications~\cite{Garcia-Bellido:2023ser,Ozsoy:2024apn,Barbon:2025wjl,Greco:2025nvv}. Note that in the simplified case where the background evolution of the scale factor and the axion field is known, the mode equations for different momenta decouple and can be solved independently, making the MbM approach straightforward to implement. This feature has also enabled the use of iterative MbM techniques to study the backreaction regime~\cite{Domcke:2020zez}.

As an alternative to the momentum-space treatment of gauge-field modes, one may work entirely in position space. In this approach, the vacuum expectation values $\langle \bm{E}^2\rangle$, $\langle \bm{B}^2\rangle$, and $\langle \bm{E}\cdot\bm{B}\rangle$ are treated as independent dynamical quantities. Their equations of motion can be derived directly from the Maxwell equations~\eqref{Maxwell_1}--\eqref{Maxwell_2}. However, a difficulty immediately arises: these equations involve new bilinear gauge-field quantities containing spatial derivatives in the form of curls, such as $\langle \bm{E}\cdot \operatorname{rot}\bm{E}\rangle$ and $\langle \bm{B}\cdot \operatorname{rot}\bm{B}\rangle$. Deriving equations of motion for these quantities, in turn, generates additional terms involving higher powers of curls.
As a result, the gauge field is described by a (in principle, infinite) hierarchy of scalar functions, given by vacuum expectation values of products of gauge fields with an increasing number of curls acting on them. For practical  applications, however, one can truncate this hierarchy at a finite order while keeping the accuracy of the results under control. This approach is known in the literature as \textit{the gradient-expansion formalism (GEF)}. It was introduced in Refs.~\cite{Sobol:2019xls,Sobol:2020lec,Gorbar:2021rlt} and has since been widely employed to study various phenomenological aspects of axion inflation and other models with inflationary gauge-field production~\cite{Gorbar:2021zlr,Durrer:2023rhc, vonEckardstein:2023gwk,Domcke:2023tnn,vonEckardstein:2024tix,vonEckardstein:2025oic, vonEckardstein:2025elq,Lysenko:2025pse,Lysenko:2025sdo}. Recently, a new Python package, the \texttt{GEF Factory}, which provides a framework for implementing and using the GEF, was released by one of the authors of the present study~\cite{vonEckardstein:2025jug}. Part of the results presented in this work have been obtained using this package.

In particular, we employ the GEF to investigate the dynamics of axion inflation across different evolutionary regimes. Since the GEF has already been discussed extensively in the literature and is used here in its existing form as a tool for numerical computations, we summarize the main equations underlying this formalism in Appendix~\ref{app:GEF} and now directly proceed to applications.


\section{Toy model: de Sitter, constant potential slope, no axion gradients}
\label{sec:const_back}


\subsection{Stationary solution and perturbations}
\label{subsec:stationary-sol}

Let us begin our analysis from a simple case of an axion rolling with a constant velocity along the slope of its potential in pure de Sitter space. Physically, this can be realized in a potential with a constant slope, $V(\phi) = V_0 - \kappa (\phi-\phi_0)$ with $\kappa=\text{const}$ and $\kappa \Delta\phi\ll V_0$, where $\Delta\phi$ in the last inequality is the typical inflaton excursion relevant for studying the gauge-field production. Although this Ansatz is taken for simplicity, it is a reasonable approximation for most of the flat inflaton potentials during a few $e$-foldings deep in the SR regime. In the absence of the gauge field, this potential admits a solution with constant inflaton velocity,
\begin{equation}
    \dot{\phi}\Big\vert_{\mathrm{no\  GF}} = -\frac{V'}{3H}=\frac{\kappa}{3H}=\text{const}\,.
\end{equation}
We will see below that, also in the presence of the gauge field, the solution with a constant rolling velocity is an exact solution of the equations of motion. Therefore, we are now interested in finding solutions with $\xi=\text{const}$.

For this, we take the mode equation \eqref{eq:mode-equation-physical-time} and use $\xi=\overline{\xi}=\text{const}$. Switching to conformal time, which is $\eta=-1/(aH)$ for de Sitter space, we obtain the following equation of motion:
\begin{equation}
\label{eq:mode-equation-constant-xi}
    \frac{\partial^2 A_\lambda(\eta, k)}{\partial\eta^2}+\bigg(k^2 +\frac{2\lambda k \overline{\xi}}{\eta}\bigg)A_\lambda(\eta, k)=0\, ,
\end{equation}
which has the form of the Whittaker equation~\cite{AbramowitzStegun:1964handbook}. Its solution satisfying the Bunch--Davies boundary condition deep inside the horizon is expressed in terms of the Whittaker $W$ function. After reverting back to physical time $t$, it has the form
\begin{equation}
\label{eq:solution-Whittaker}
    A_\lambda(t,k)=\frac{e^{\lambda \pi\overline{\xi}/2}}{\sqrt{2k}}W_{-i\lambda\overline{\xi},\,\frac{1}{2}}\Big(-2i\frac{k}{a(t)H}\Big)\, .
\end{equation}
Further, we use this solution to compute the correlator on the right-hand side of the Klein--Gordon equation~\eqref{eq:Klein-Gordon-final}:
\begin{equation}
\label{eq:EB-function-H-xi}
    \langle\bm{E}\cdot\bm{B}\rangle = - H^4\,\overline{g}_0(\overline{\xi})
\end{equation}
with%
\footnote{Here, the integration variable is $x=k/(aH)$, and we used the expression for the derivative of the Whittaker function $dW_{\sigma,\mu}(y)/d\ln y = (y/2-\sigma)W_{\sigma,\mu}(y)-W_{1+\sigma,\mu}(y)$; see, e.g., Eq.~(13.4.33) in Ref.~\cite{AbramowitzStegun:1964handbook}.}
\begin{equation}
    \overline{g}_0(\overline{\xi})=\frac{1}{4\pi^2} \int\limits_0^{2|\overline{\xi}|}dx\,\sum\limits_{\lambda=\pm}\lambda x^2 e^{\lambda\pi\overline{\xi}} \operatorname{Re}\Big[W_{i\lambda\overline{\xi},\,\frac{1}{2}}(2ix)W_{1-i\lambda\overline{\xi},\,\frac{1}{2}}(-2ix)\Big]\, .
\end{equation}
Since we consider pure de Sitter space, the Friedmann equation~\eqref{eq:Friedmann-final} is excluded from the analysis. Expressing $\dot{\phi}=2HM_{\mathrm{P}}\overline{\xi}/\beta$, we obtain from the Klein--Gordon equation~\eqref{eq:Klein-Gordon-final}
\begin{equation}
    \ddot{\phi} = 0 = -\frac{6H^2 M_{\mathrm{P}}}{\beta}\,\overline{\xi} - V' - \frac{\beta}{M_{\mathrm{P}}} H^4\,\overline{g}_0(\overline{\xi}) \, ,
\end{equation}
or, in a simpler form,
\begin{tcolorbox}
\begin{equation}
\label{eq:stationary-solutions}
6\,\overline{\xi} - b\,v  + b^2\,\overline{g}_0(\overline{\xi}) = 0 \vphantom{\frac{1}{2}}\,,
\end{equation}
\end{tcolorbox}
\noindent 
where $b=\beta H/M_{\mathrm{P}}$ is the axion--vector coupling in Hubble units and
\begin{equation}
\label{eq:def-tildev}
     v  = -\frac{V'}{H^3} = \frac{3\sqrt{2\epsilon_V}}{H/M_\mathrm{P}} = \frac{3}{2\pi \sqrt{\mathcal{P}_{\mathcal{R}}}}
\end{equation}
is a parameter characterizing the inflaton potential, namely the potential gradient expressed in Hubble units. Here, $\epsilon_V = (M_\mathrm{P}^2/2)(V'/V)^2$ is the first slow-roll parameter and $\mathcal{P}_{\mathcal{R}} = H^2/(8\pi^2 \epsilon_V M_\mathrm{P}^2)$ is the amplitude of the primordial scalar power spectrum generated during slow-roll inflation in the given potential. Thus, Eq.~\eqref{eq:stationary-solutions} implicitly determines the stationary solution of the equations of motion for given inflaton potential (in terms of $v$) and axion--vector coupling in Hubble units, $b$. The contour plot of the function $\overline{\xi}=\overline{\xi}(b, v)$ is shown in Fig.~\ref{fig:instability1} by black solid lines.

Further, we consider small perturbations on top of this stationary solution. Since we are working within the approximation of a homogeneous axion field, the perturbations we consider are only functions of time $t$ or the number of $e$-foldings $N = \ln a = H t$, which is another convenient dimensionless timelike variable during inflation. For the instability variable, we use an Ansatz
\begin{equation}
    \xi(N)= \overline{\xi} + \delta\xi(N)\, .
\end{equation}
As for the gauge-field bilinear quantities, such as $\langle\bm{E}\cdot\bm{B}\rangle$, which enters the Klein--Gordon equation, one cannot express them as functions of $H$ and $\xi$ as done in the case of the stationary solution; cf.\ Eq.~\eqref{eq:EB-function-H-xi}. In general, they are functionals of $\xi$ that take into account the effects of the retarded response of the gauge field to changes in $\xi$. Therefore, we must treat perturbations of the gauge-field configuration on the same footing as axion perturbations.

There are different approaches to this issue. First, one could work with the gauge-field mode functions in Fourier space and introduce the corresponding perturbation $\delta A_\lambda$ as the deviation of the full mode function from its expression~\eqref{eq:solution-Whittaker} in the case of constant $\xi$. One can then derive a closed system of equations for $\delta\xi$ and $\delta A_\lambda$ within linear perturbation theory. This approach was realized in Ref.~\cite{Peloso:2022ovc}, where the stability of the stationary solutions in the strong-backreaction regime was studied for the first time. However, to obtain the growth rate of the perturbations, one has to solve a complicated transcendental equation, which is computationally very costly. Alternatively, one can continue working with the gauge-field bilinear functions in position space and study the properties of their small perturbations. For this purpose, in Ref.~\cite{vonEckardstein:2023gwk}, some of us developed a fast and efficient tool\,---\,\textit{the linearized gradient-expansion formalism} (LGEF)\,---\,which is a version of the more general GEF approach linearized with respect to small perturbations around a stationary solution. Below, we briefly review the details of this approach.

First, we define a set of bilinear gauge-field correlators
\begin{align}
    e_n(N) &= \phantom{-}\frac{1}{a^n H^{n+4}}\langle \bm{E}\cdot\operatorname{rot}^{n}\bm{E}\rangle = \Big(\frac{k_{\mathrm{h}}}{aH}\Big)^{n+4}\mathcal{F}_E^{(n)}\, ,\\
    g_n(N) &=- \frac{1}{a^n H^{n+4}}\langle \bm{E}\cdot\operatorname{rot}^{n}\bm{B}\rangle= \Big(\frac{k_{\mathrm{h}}}{aH}\Big)^{n+4}\mathcal{F}_G^{(n)}\, ,\\
    b_n(N) &= \phantom{-}\frac{1}{a^n H^{n+4}}\langle \bm{B}\cdot\operatorname{rot}^{n}\bm{B}\rangle= \Big(\frac{k_{\mathrm{h}}}{aH}\Big)^{n+4}\mathcal{F}_B^{(n)}\,,
\end{align}
where $n$ is a non-negative integer and $\mathcal{F}_X^{(n)}$ are the bilinear functions of the full GEF given in Appendix~\ref{app:GEF}.

For the stationary solution with $\xi=\overline{\xi}=\text{const}$, these bilinear functions can be expressed as functions of $\overline{\xi}$ only, by using the exact mode solutions in Eq.~\eqref{eq:solution-Whittaker}:
\begin{align}
    \overline{e}_n(\overline{\xi})&=\frac{1}{4\pi^2} \int\limits_0^{2|\overline{\xi}|}dx\,\sum\limits_{\lambda=\pm}\lambda^{n} x^{n+1} e^{\lambda\pi\overline{\xi}} \Big|(x-\lambda \overline{\xi})W_{-i\lambda\overline{\xi},\,\frac{1}{2}}(-2ix)-iW_{1-i\lambda\overline{\xi},\,\frac{1}{2}}(-2ix)\Big|^2\, ,\label{eq:e-n-function}\\
    \overline{g}_n(\overline{\xi})&=\frac{1}{4\pi^2} \int\limits_0^{2|\overline{\xi}|}dx\,\sum\limits_{\lambda=\pm}\lambda^{n+1} x^{n+2} e^{\lambda\pi\overline{\xi}} \operatorname{Re}\Big[W_{i\lambda\overline{\xi},\,\frac{1}{2}}(2ix)W_{1-i\lambda\overline{\xi},\,\frac{1}{2}}(-2ix)\Big]\, ,\label{eq:g-n-function}\\
    \overline{b}_n(\overline{\xi})&=\frac{1}{4\pi^2} \int\limits_0^{2|\overline{\xi}|}dx\,\sum\limits_{\lambda=\pm}\lambda^n x^{n+3} e^{\lambda\pi\overline{\xi}} \Big|W_{-i\lambda\overline{\xi},\,\frac{1}{2}}(-2ix)\Big|^2\, .\label{eq:b-n-function}
\end{align}
Then, the linear perturbations are introduced as follows:
\begin{align}
    \delta e_n(N) = e_n(N) - \overline{e}_n(\overline{\xi})\, ,\\
    \delta g_n(N) = g_n(N) - \overline{g}_n(\overline{\xi})\, ,\\
    \delta b_n(N) = b_n(N) - \overline{b}_n(\overline{\xi})\, .
\end{align}
Together with $\delta\xi$, they constitute the full set of perturbations in our problem. To obtain the equations of motion for these perturbations in the linear approximation, we use the Klein--Gordon equation~\eqref{eq:Klein-Gordon-final} together with the GEF equations from Appendix~\ref{app:GEF}. The final result reads
\begin{align}
\label{KGF-xi-delta}
& \frac{d\delta \xi}{dN}+3\delta \xi =-\frac{b^2}{2} \delta  g_0 \,, \\
\label{E_n_const-delta}
& \frac{d\delta  e_n}{dN}+(n+4)\delta  e_n-4\overline{\xi}\, \delta  g_n-4\overline{g}_n(\overline{\xi})\,\delta \xi+2\delta  g_{n+1}= \delta  S^{(e)}_n \,, \\
\label{G_n_const-delta}
& \frac{d\delta  g_n}{dN}+(n+4)\delta  g_n-2\overline{\xi} \delta  b_n-2\overline{b}_n(\overline{\xi})\,\delta  \xi+\delta  b_{n+1}-\delta  e_{n+1} = \delta  S^{(g)}_n \,, \\
\label{B_n_const-delta}
& \frac{d\delta  b_n}{dN}+(n+4)\delta  b_n-2\delta  g_{n+1} = \delta  S^{(b)}_n \,, 
\end{align}
where the sources have the form
\begin{align}
\delta  S^{(e)}_n & = \frac{(2\overline{\xi})^{n+4}}{4\pi^2}\sum\limits_{\lambda=\pm}\lambda^n \bigg\{\frac{E_\lambda(\overline{\xi})}{\overline{\xi}}\left[(n+4)\delta  \xi+\frac{d\delta \xi}{dN}\right]+\frac{dE_\lambda(\overline{\xi})}{d\overline{\xi}}\delta  \xi\bigg\} \,, \\
\delta  S^{(g)}_n & = \frac{(2\overline{\xi})^{n+4}}{4\pi^2}\sum\limits_{\lambda=\pm}\lambda^{n+1} \bigg\{\frac{G_\lambda(\overline{\xi})}{\overline{\xi}}\left[(n+4)\delta  \xi+\frac{d\delta \xi}{dN}\right]+\frac{dG_\lambda(\overline{\xi})}{d\overline{\xi}}\delta  \xi\bigg\} \,, \\
\delta  S^{(b)}_n & = \frac{(2\overline{\xi})^{n+4}}{4\pi^2}\sum\limits_{\lambda=\pm}\lambda^n \bigg\{\frac{B_\lambda(\overline{\xi})}{\overline{\xi}}\left[(n+4)\delta  \xi+\frac{d\delta \xi}{dN}\right]+\frac{dB_\lambda(\overline{\xi})}{d\overline{\xi}}\delta  \xi\bigg\} \,, 
\end{align}
and the functions $E_\lambda$, $G_\lambda$, and $B_\lambda$ are defined in Eqs.~\eqref{eq: Source Terms} in Appendix~\ref{app:GEF}. This system is infinite in principle and, in order to use it in practice, one has to truncate it at some order $n_\mathrm{cut}$. The simplest way to do this is to assume that, for all orders larger than $n_\mathrm{cut}$, the bilinear functions exactly coincide with the background values of the stationary solution,
\begin{equation}
\label{LGEF-truncation}
    \delta e_n = \delta g_n = \delta b_n = 0 \qquad \text{for} \qquad n > n_\mathrm{cut}.
\end{equation}
One has to choose $n_{\mathrm{cut}}$ in such a way that further increasing it does not change the numerical solutions within the desired accuracy. In the next subsection, we apply the LGEF approach to compute the spectrum of Lyapunov exponents of our system over a wide range of model parameters.


\subsection{Lyapunov exponents}
\label{subsec:Lyapunov}

The set of equations of motion for perturbations in the LGEF approach derived in the previous subsection forms a linear, homogeneous system of first-order ordinary differential equations with constant coefficients,
\begin{equation}
\label{eq:LGEF-matrix-form}
    \frac{d}{dN}\vec{x}(N) = \underline{\mathbf{A}} \vec{x}(N)\, ,
\end{equation}
where
\begin{equation}
  \vec{x}\equiv
  \left[
    \begin{array}{c}
      \delta\xi \\
      \left\{
        \begin{array}{c}
          \delta e_n \\
          \delta g_n \\
          \delta b_n
        \end{array}
      \right\}
    \end{array}
  \right]
\end{equation}
is a $(3n_{\rm cut}+4)$-dimensional vector of perturbations.
The curly brackets $\{\dots\}$ denote a set of variables labeled by $n\in[0,\,n_{\mathrm{cut}}]$. The system matrix $\underline{\mathbf{A}}$ is uniquely determined by the linearized equations \eqref{KGF-xi-delta}--\eqref{B_n_const-delta}.

To solve this system, we make the standard Ansatz
\begin{equation}
  \vec{x}(N) = \vec{x}_0 \, e^{\zeta N}\, ,
\end{equation}
where $\vec{x}_0$ is a constant vector and $\zeta$ is, in general, a complex number.
Substituting this Ansatz into Eq.~\eqref{eq:LGEF-matrix-form}, the problem reduces to the linear eigenvalue problem
\begin{equation}
    (\underline{\mathbf{A}} - \zeta \underline{\mathbb{1}})\vec{x}_0 = \vec{0}\, .
\end{equation}
The eigenvalues $\zeta$ are the Lyapunov exponents of the linearized system and fully determine the stability properties of the stationary solution $\xi=\overline{\xi}=\mathrm{const}$.
Importantly, their computation is a purely algebraic task\,---\,it does not require solving differential equations.

For each point in the two-dimensional parameter space $(v,\,b)$, we first determine the corresponding stationary value $\overline{\xi}$, construct the matrix $\underline{\mathbf{A}}$, and compute its eigenvalues. These eigenvalues are ordered according to their real part. The eigenvalue%
\footnote{Complex eigenvalues always appear in conjugate pairs as $\underline{\mathbf{A}}$ is a real matrix. For definiteness, we count only one eigenvalue from each pair, choosing the one with positive imaginary part.}
with the largest real part, denoted $\zeta_1$, controls the (in)stability of the stationary solution.

\begin{figure}
\centering
\includegraphics[width=0.73\linewidth]{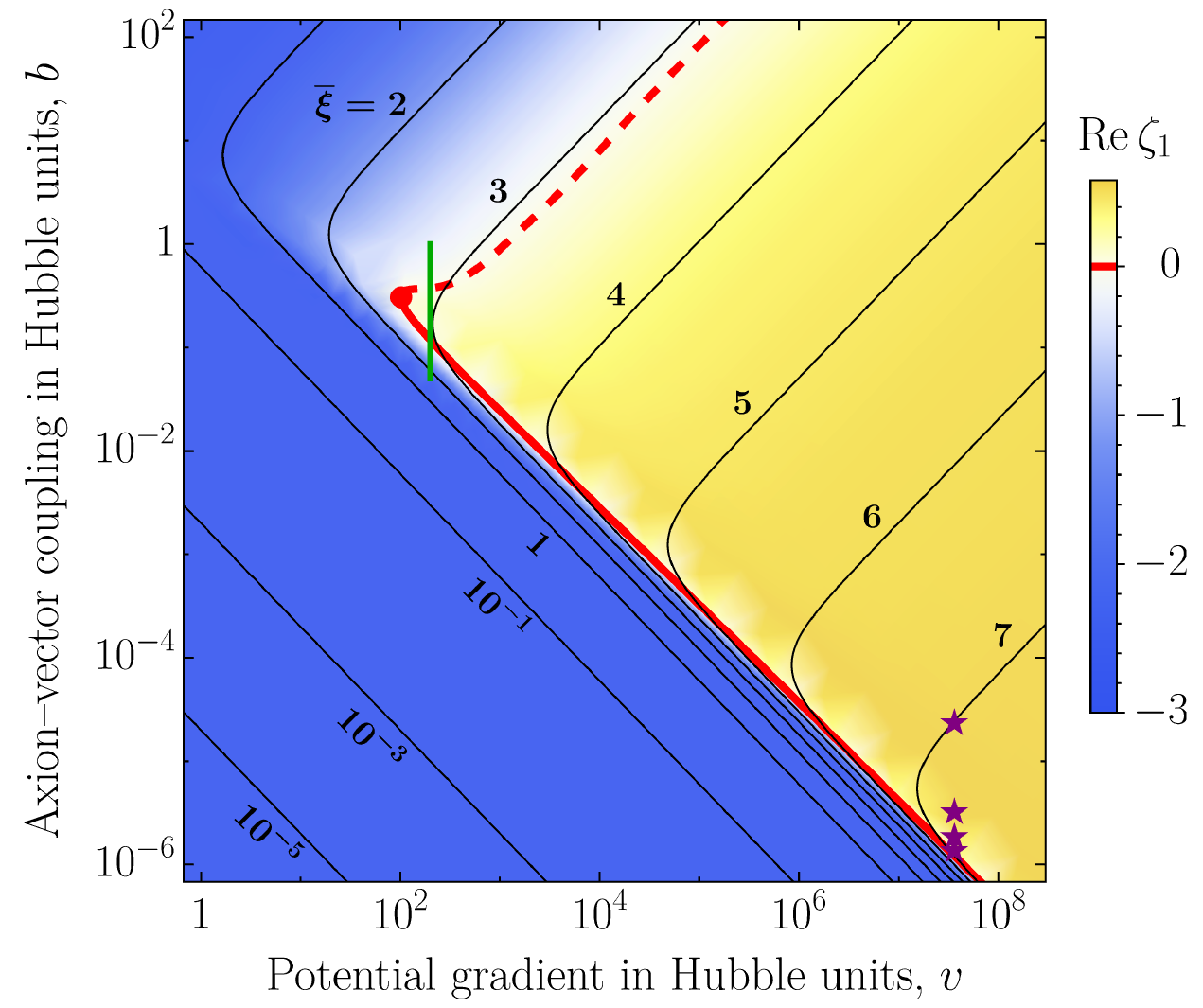}
\caption{Stability diagram in the LGEF approach with $n_\mathrm{cut}=70$. The background color shows the largest real part among all Lyapunov exponents, $\operatorname{Re}\zeta_1$, as a function of the axion--vector coupling in Hubble units, $b=\beta H/M_p$, and the potential gradient in Hubble units, $v = -V'/H^3$. The solid and dashed parts of the red line correspond to $\operatorname{Re}\zeta_1=0$ and mark the boundary of the unstable region when approached from the ordinary SR regime and from the stable backreaction regime, respectively. The red dot at the merger of the solid and dashed red curves denotes the codimension-2 bifurcation point (see the discussion in Sec.~\ref{subsec:bifurcation}). The solid black lines show contours of constant $\overline{\xi}$. The solid green segment denotes the crossing of the instability band for which the bifurcation diagram is constructed in Sec.~\ref{subsec:bifurcation} (cf. Fig.~\ref{fig:bifurcation}).
The purple stars indicate the parameter points studied in the strong-backreaction regime in Sec.~\ref{subsec:bursting} (cf. Fig.~\ref{fig:bursting}). }
\label{fig:instability1}
\end{figure}

The real part of $\zeta_1$ is shown in Fig.~\ref{fig:instability1} as a function of the axion--vector coupling $b$ and the gradient parameter $v$. Negative values of $\operatorname{Re}\zeta_1$ (shaded in blue) correspond to a stationary solution that is stable against small perturbations, while positive values (shaded in yellow) indicate instability. The red contour marks the boundary $\operatorname{Re}\zeta_1=0$, separating these two regimes.

The black contour lines indicate surfaces of constant $\overline{\xi}$, ranging from $10^{-5}$ to $7$. In the lower-left corner of the plot, corresponding to $\overline{\xi}<0.1$, gauge-field production is negligible and inflation proceeds in the standard SR regime. In this limit, the last term in Eq.~\eqref{KGF-xi-delta} can be neglected, yielding the Lyapunov exponent $\zeta_1=-3$, which is clearly visible in the plot. Perturbations are simply redshifted away and not further sourced by the interaction with the gauge field.

An interesting feature of Fig.~\ref{fig:instability1} is that, for fixed $v>102$ (to the right from the red dot in Fig.~\ref{fig:instability1}), increasing the axion--vector coupling first drives the system into the unstable region (crossing the solid red line), but at sufficiently large coupling the system re-enters a stable regime (crossing the dashed red line). This second stable region does not correspond to slow-roll inflation; instead, the produced gauge field becomes strong and its effective friction term plays a dominant role in the Klein--Gordon equation~\eqref{KGF-2}.

\begin{figure}
\centering
\includegraphics[width=0.68\linewidth]{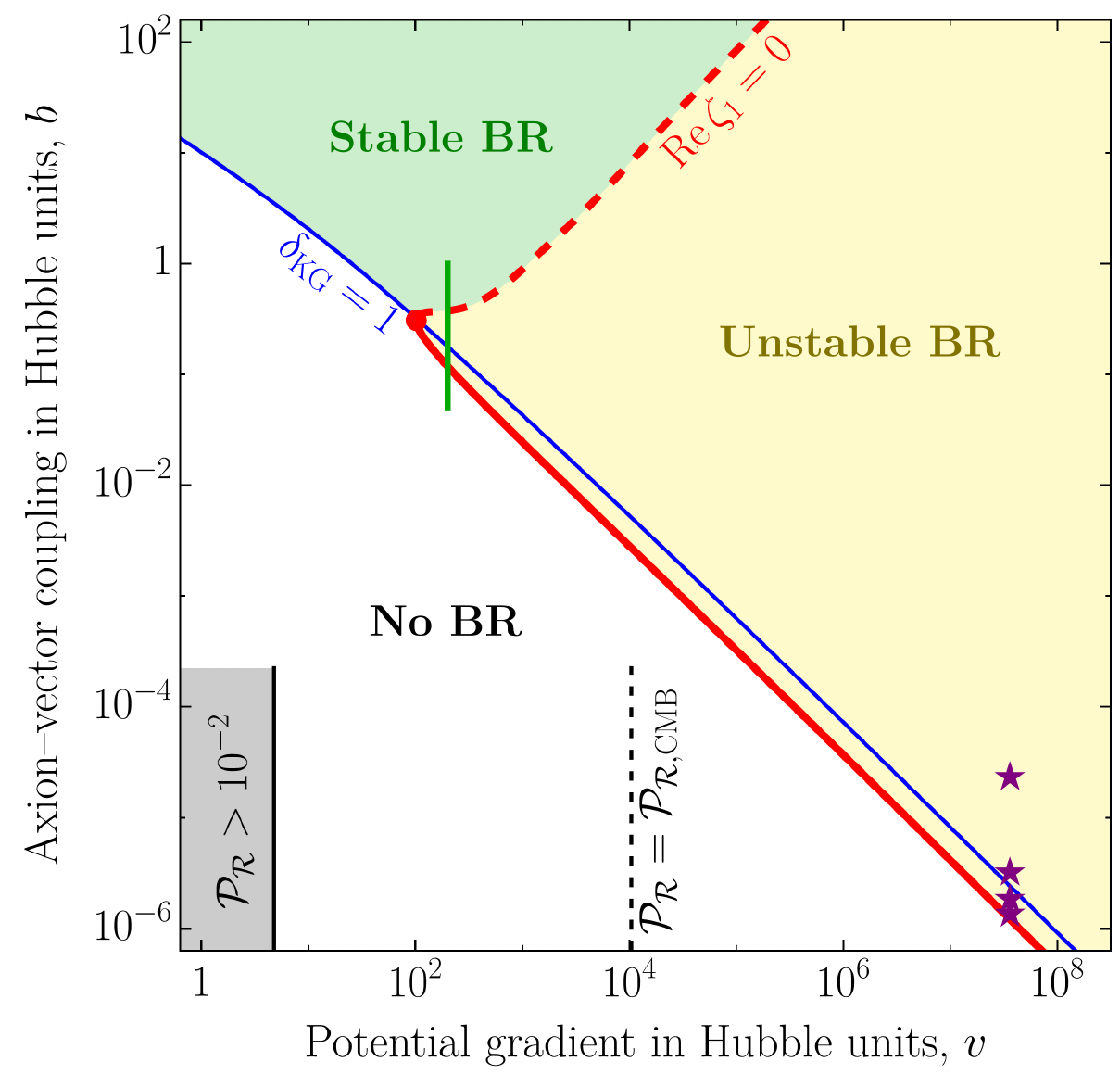}
\caption{Parameter-space regions corresponding to negligible backreaction (white), unstable backreaction (yellow shaded), and stable backreaction (green shaded). The blue solid line indicates the conventional backreaction criterion $\delta_{\mathrm{KG}}=1$.
The vertical dashed black line marks the value of $v$ fixed by the CMB normalization. The gray region corresponds to excessively large vacuum scalar perturbations, $\mathcal{P}_{\mathcal{R}}>10^{-2}$, where the approximation of a homogeneous inflaton field breaks down even in the absence of gauge fields. The rest of the notation is the same as in Fig.~\ref{fig:instability1}.}
\label{fig:backreaction}
\end{figure}

To quantify the importance of the gauge-field friction in the Klein--Gordon equation~\eqref{KGF-2}, we define the backreaction parameter~\cite{Jimenez:2017cdr}
\begin{equation}
\label{eq:delta-KG-def}
    \delta_{\mathrm{KG}} =
    \left|
      \frac{
        \frac{\beta}{M_\mathrm{P}}
        \langle \bm{E}\cdot\bm{B}\rangle
      }{
        3H\dot{\phi}
      }
    \right|\, .
\end{equation}
Figure~\ref{fig:backreaction} shows, in less details, the same parameter space as in Fig.~\ref{fig:instability1}. In addition, the blue solid line corresponds to $\delta_{\mathrm{KG}}=1$ and above this curve $\delta_{\mathrm{KG}}>1$. One can clearly distinguish the white region where the system possesses a stable solution (which reduces to the pure slow-roll attractor far below from the blue line). 

Furthermore, in Fig.~\ref{fig:backreaction}, there is another region (shaded in green) where the stationary solution $\overline{\xi}(b,v)$ is stable and, at the same time, where the gauge field strongly backreacts onto the inflaton dynamics (since $\delta_{\mathrm{KG}}>1$ there). In this region, purely time-dependent and spatially homogeneous perturbations remain bounded, even though backreaction is strong. Within our semi-analytical framework, we cannot rigorously demonstrate that spatial inhomogeneities are also suppressed. Nevertheless, we argue that this is likely the case. A purely time-dependent perturbation corresponds to the $k=0$ limit of a general spacetime-dependent perturbation. On general grounds, this behavior should extend to superhorizon modes $k/(aH)\ll1$, which are observationally indistinguishable from $k=0$ modes for subhorizon observers.
Moreover, for $H=\mathrm{const}$ and $\xi=\mathrm{const}$, the induced scalar power spectrum (sourced by the gauge field) is flat on superhorizon scales~\cite{Barnaby:2011vw,Barnaby:2010vf,Barnaby:2011qe,Durrer:2024ibi},
$\mathcal{P}_{\mathcal{R}}^{\mathrm{ind}}=\mathrm{const}$. If this spectrum is small for the longest wavelengths, it should also remain small for modes crossing the horizon later during inflation. In other words, if $\mathcal{P}_{\mathcal{R}}^{\mathrm{ind}}(k=0)$ experiences no instability, we expect that the same is true for $\mathcal{P}_{\mathcal{R}}^{\mathrm{ind}}(k \ll aH)$ and hence, ultimately, for $\mathcal{P}_{\mathcal{R}}^{\mathrm{ind}}(k \gg aH)$. We therefore expect the homogeneous inflationary solution to act as a stable attractor in this region. 

Another argument in support of this conclusion is that, throughout the entire \textit{stable-backreaction (SB) region}, the instability variable satisfies $\overline{\xi}\lesssim 3.14$, which lies below the non-perturbativity bound $\xi_{\text{n-p}}\approx 4.6$ reported in Refs.~\cite{Ferreira:2015omg,Peloso:2016gqs}. Therefore, within this newly identified stable region, scalar perturbations (particularly axion inhomogeneities) are expected to remain under perturbative control. 
The arguments presented in the last two paragraphs are heuristic rather than rigorous, and a complete treatment of perturbations in this parameter range is deferred to future work.

To the best of our knowledge, this stable-backreaction region has not been reported previously in the literature. Indeed, previous semi-analytical studies of the strong-backreaction regime in Abelian axion inflation~\cite{Domcke:2020zez,Cheng:2015oqa,Notari:2016npn,Gorbar:2021rlt,Peloso:2022ovc,vonEckardstein:2023gwk,vonEckardstein:2025oic} have shown that the solution exhibits oscillatory behavior and deviates substantially from the smooth Anber--Sorbo trajectory. Lattice simulations~\cite{Caravano:2022epk,Figueroa:2023oxc,Sharma:2024nfu,Figueroa:2024rkr,Jamieson:2025ngu,Lizarraga:2025aiw} further demonstrated that axion spatial inhomogeneities are strongly enhanced in this regime.
However, all such studies focused on parameter values lying within the yellow shaded region of Fig.~\ref{fig:backreaction}, which we identify as the \emph{unstable-backreaction (UB) region}. For a detailed comparison of the parameter space regions studied in the literature, see Appendix~\ref{app:comp-previous}. Notably, the SB regime can be realized only for sufficiently strong axion--vector coupling expressed in Hubble units, $b>\mathcal{O}(1)$. To our knowledge, none of the previous studies have explored this region of parameter space; see Fig.~\ref{fig:models} in the Appendix.

Finally, we note the existence of parameter values for which $\delta_{\mathrm{KG}}<1$, yet the stationary solution is already unstable, as seen in the yellow-shaded band below the blue $\delta_{\mathrm{KG}}=1$ line in Fig.~\ref{fig:backreaction}.
In this regime, the gauge-field-induced friction term can be as small as $\sim 5\,\%$ of the Hubble friction, and the background still follows a slow-roll trajectory. Nevertheless, the feedback from gauge-field production is sufficient to destabilize the slow-roll attractor.

This observation motivates us to introduce a new backreaction criterion for $v>102$:
\begin{tcolorbox}
\begin{equation}
\label{eq:criterion-BR-new}
    \operatorname{Re}\zeta_1 \geq 0\, ,\vphantom{\frac{1}{2}}
\end{equation}
\end{tcolorbox}

\noindent
which is more restrictive than the conventional condition $\delta_{\mathrm{KG}}\geq 1$. Moreover, Eq.~\eqref{eq:criterion-BR-new} provides a different interpretation of gauge-field backreaction. The widely used criterion $\delta_{\mathrm{KG}} \geq 1$ indicates that the gauge field has become sufficiently strong for its contribution to the inflationary dynamics to be comparable to that of the inflaton potential gradient. By contrast, the new criterion implies that the presence of the gauge field, although possibly negligible at the moment when the criterion is satisfied for the first time, will eventually have a significant impact on the dynamics. In the next subsection, we present a detailed analysis of this new criterion and derive simple fit functions for the corresponding threshold values of the parameters. In Sec.~\ref{sec:realistic}, we then apply this dynamical criterion to two realistic inflationary models and show that the slow-roll trajectory is not disrupted immediately, but rather within $\mathcal{O}(1)$ $e$-foldings after the condition~\eqref{eq:criterion-BR-new} is met.


\subsection{Backreaction criterion}
\label{subsec:criterion}

\begin{figure}
\centering
\includegraphics[height=4.85cm]{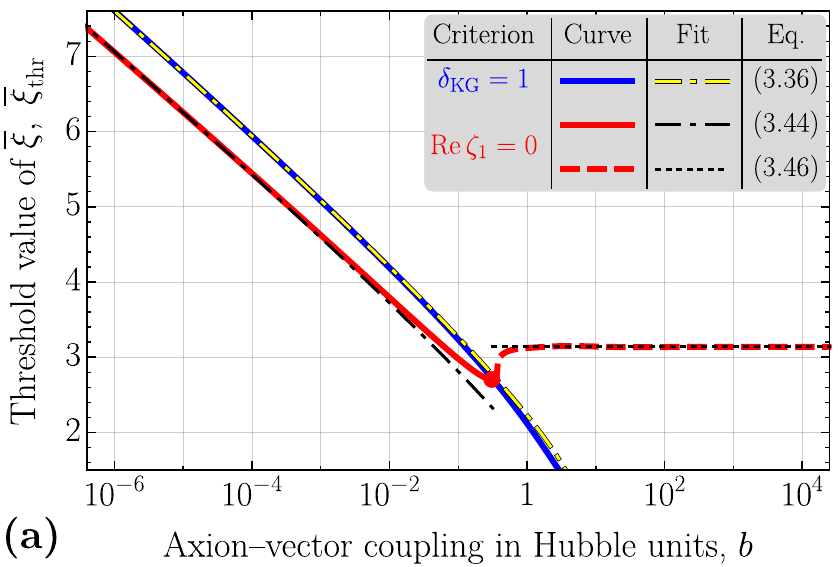}\hspace{3mm}
\includegraphics[height=4.85cm]{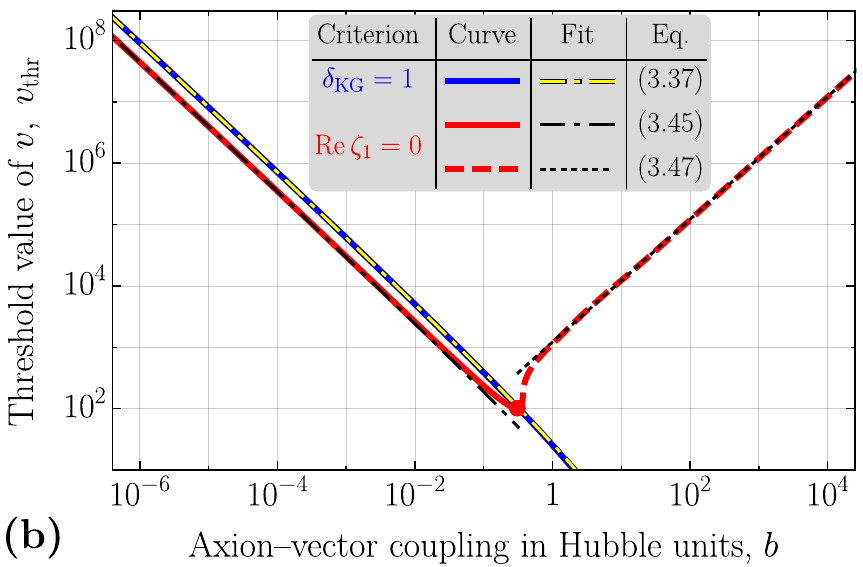}
\caption{Panel~(a): Threshold value of the instability variable, $\overline{\xi}_\mathrm{thr}$, at which the system enters the strong-backreaction regime, shown as a function of the axion--vector coupling $b=\beta H/M_{\mathrm{P}}$. 
The blue and red curves have the same meaning as in Fig.~\ref{fig:backreaction}. 
The blue solid line corresponds to the old backreaction criterion, $\delta_{\mathrm{KG}}=1$, defined by the relative magnitude of the gauge-field contribution in the Klein--Gordon equation. 
The red solid line represents the new criterion, $\operatorname{Re}\zeta_1 = 0$, based on the stability analysis of linear perturbations. 
The red dashed curve also corresponds to $\operatorname{Re}\zeta_1 = 0$, but occurring within the strong-backreaction regime; it therefore marks the boundary between the SB and UB regimes. 
The yellow dashed-dotted line shows the approximate expression given in Eq.~\eqref{eq:fit-xi-old-BR}. 
The black dashed-dotted line corresponds to the fit function in Eq.~\eqref{eq:fit-xi-new-BR}, while the black dotted line represents the fit in Eq.~\eqref{eq:fit-xi-stable-BR}.
Panel~(b): Threshold value of the potential-gradient parameter, $v_\mathrm{thr}$, as a function of $b$. 
The blue curve and the solid and dashed red curves have the same interpretation as in panel~(a). 
The yellow dashed-dotted line shows the approximate relation given in Eq.~\eqref{eq:fit-v-old-BR}, the black dashed-dotted line corresponds to the fit function in Eq.~\eqref{eq:fit-v-new-BR}, and the black dotted line represents the fit in Eq.~\eqref{eq:fit-v-stable-BR}.
\label{fig:thresholds}}
\end{figure}

Let us now examine the backreaction criteria introduced above in greater detail. 
In what follows, we analyze separately the old criterion based on the parameter $\delta_{\mathrm{KG}}$ (corresponding to the blue solid line in Fig.~\ref{fig:backreaction}) and the new criterion based on the Lyapunov exponents of the linearized system (represented by the solid and dashed red lines in Fig.~\ref{fig:backreaction}). 
For each of these three curves, we determine the threshold values of the instability variable, $\bar{\xi}_\mathrm{thr}$, and of the potential-gradient parameter, $v_\mathrm{thr}$, as functions of the axion--vector coupling $b$. This analysis is performed numerically, and we additionally derive analytic fit functions that accurately reproduce the numerical results and are straightforward to implement in practical applications. The exact numerical curves together with their corresponding fits are shown in Fig.~\ref{fig:thresholds}.

\subsubsection*{Old criterion: $\boldsymbol{\delta_{\mathrm{KG}}=1}$}

Given the definition of the parameter $\delta_{\mathrm{KG}}$ in Eq.~\eqref{eq:delta-KG-def}, the old backreaction criterion $\delta_{\mathrm{KG}}=1$, together with the stationary Klein--Gordon equation~\eqref{eq:stationary-solutions}, implies
\begin{equation}
\label{eq:b-of-xi-old}
    b=\sqrt{\frac{6\overline{\xi}}{\overline{g}_0(\overline{\xi})}}\, , 
    \qquad 
    v=\frac{12\overline{\xi}}{b}
    =2\sqrt{6\overline{\xi}\,\overline{g}_0(\overline{\xi})}\, ,
\end{equation}
which define the blue solid curves in Figs.~\ref{fig:backreaction} and~\ref{fig:thresholds} in parametric form. The curve in Fig.~\ref{fig:backreaction} has a simple and instructive interpretation: it consists of the points in the $(v,b)$ plane at which $\overline{\xi}$ attains its maximal value as a function of $b$ for fixed $v$.
Indeed, differentiating Eq.~\eqref{eq:stationary-solutions} with respect to $b$ at fixed $v$ and imposing $d\overline{\xi}/db=0$, we immediately obtain $v = 2b\,\overline{g}_0(\overline{\xi})$.
Combining this relation with Eq.~\eqref{eq:stationary-solutions}, one recovers Eq.~\eqref{eq:b-of-xi-old} above.

In principle, Eq.~\eqref{eq:b-of-xi-old} already contains all the information required for practical applications, since the function $\overline{g}_0(\overline{\xi})$ is given in integral form in Eq.~\eqref{eq:g-n-function}. 
However, for convenience, we derive simpler fit expressions that avoid the explicit evaluation of integrals.
To this end, we consider the large-$\overline{\xi}$ limit and use the asymptotic form~\cite{Jimenez:2017cdr,Sobol:2019xls}
\begin{equation}
\label{eq:g0-approx}
    \overline{g}_0(\overline{\xi}) \simeq 
    C\,\frac{e^{2\pi\overline{\xi}}}{\overline{\xi}^4}\, ,
\end{equation}
where $C=9/(1120\pi^3)\approx 2.59\times 10^{-4}$. 
Substituting this expression into the first relation in Eq.~\eqref{eq:b-of-xi-old}, we obtain
\begin{equation}
    b\simeq\sqrt{\frac{6}{C}}\,
    \overline{\xi}^{5/2} e^{-\pi \overline{\xi}}
    \approx 152\,\overline{\xi}^{5/2} e^{-\pi \overline{\xi}} \,.
\end{equation}
Solving this equation for $\overline{\xi}$ yields the following expression for the threshold value of the instability parameter as a function of the axion--vector coupling $b$:
\begin{equation}
\label{eq:fit-xi-old-BR}
    \overline{\xi}_\mathrm{thr}^{\mathrm{old}}(b)
    \approx -\frac{5}{2\pi}
    \mathcal{W}_{-1}\!\left[
    -\frac{2\pi}{5}
    \left(\frac{b}{152}\right)^{2/5}
    \right] ,
\end{equation}
where $\mathcal{W}_{-1}(x)$ denotes the negative branch of the Lambert $W$ function, defined by $\mathcal{W}e^{\mathcal{W}}=x$ for $x\in[-e^{-1},0)$. 

Using the second relation in Eq.~\eqref{eq:b-of-xi-old}, we further obtain the approximate expression for the threshold value of the potential-gradient parameter $v$:
\begin{equation}
\label{eq:fit-v-old-BR}
    v_\mathrm{thr}^\mathrm{old}(b)
    =\frac{12\,\overline{\xi}_\mathrm{thr}^{\mathrm{old}}(b)}{b}
    \approx -\frac{30}{\pi b}
    \mathcal{W}_{-1}\!\left[
    -\frac{2\pi}{5}
    \left(\frac{b}{152}\right)^{2/5}
    \right] .
\end{equation}
Equations~\eqref{eq:fit-xi-old-BR} and~\eqref{eq:fit-v-old-BR} provide an excellent approximation to the blue curves in Fig.~\ref{fig:thresholds}(a) and~\ref{fig:thresholds}(b), respectively, for $b<1$. 
In the figure, these approximations are shown by the yellow dashed-dotted lines in the corresponding panels. For $b\gtrsim 1$, the accuracy of the fit deteriorates, since its derivation relies on the large-$\overline{\xi}$ approximation given in Eq.~\eqref{eq:g0-approx}.

\subsubsection*{New criterion: $\boldsymbol{\operatorname{Re}\zeta_1=0}$ at the SR--UB boundary}

We now turn to the red solid line in Figs.~\ref{fig:instability1} and~\ref{fig:backreaction}. 
For small $b\ll 1$, Fig.~\ref{fig:instability1} shows that this contour passes through a region where the behavior of $\overline{\xi}(b)$ changes qualitatively: an initially rapid growth with $b$ transitions to a much slower increase and eventually to a mild decrease. 
This observation suggests that an analytic approximation to the red curve can be obtained by studying the logarithmic derivative $d\ln\overline{\xi}/d\ln b$ and comparing the respective contributions from the slow-roll inflaton dynamics and the gauge-field backreaction. 
When these two contributions become comparable, one expects the onset of backreaction according to our new criterion.

We again employ Eq.~\eqref{eq:stationary-solutions} and solve it for $b$:
\begin{equation}
\label{eq:b-of-xi-v-full}
    b=\frac{12\overline{\xi}}{v+\sqrt{v^2-24\,\overline{\xi}\,\overline{g}_0(\overline{\xi})}}\, .
\end{equation}
For $b\ll 1$, the red solid line in Fig.~\ref{fig:backreaction} lies well below the blue one, implying that along this curve the gauge-field contribution to the Klein--Gordon equation remains subdominant compared to the slow-roll terms. 
We therefore expand Eq.~\eqref{eq:b-of-xi-v-full} to first nontrivial order in $\overline{g}_0(\overline{\xi})$. 
Taking the logarithm yields
\begin{equation}
\label{eq:b-of-xi-v-appr}
    \ln b \simeq \ln\!\left(\frac{6\overline{\xi}}{v}\right)
    + \frac{6}{v^2}\,\overline{\xi}\,\overline{g}_0(\overline{\xi})\, .
\end{equation}
The first term represents the slow-roll inflaton contribution, while the second term is the correction induced by the gauge field. 
Taking the logarithmic derivative with respect to $\overline{\xi}$ gives
\begin{equation}
    \frac{d\ln b}{d\ln\overline{\xi}}
    \simeq 1 + \frac{6}{v^2}
    \Big[
    \overline{\xi}\,\overline{g}_0(\overline{\xi})
    + \overline{\xi}^2\,\overline{g}_0'(\overline{\xi})
    \Big] .
\end{equation}
Using Eq.~\eqref{eq:stationary-solutions} once more and neglecting the gauge-field contribution, we approximate $v\simeq 6\overline{\xi}/b$, which leads to
\begin{equation}
    \frac{d\ln b}{d\ln\overline{\xi}}
    \simeq 1 + \frac{b^2}{6}
    \left[
    \frac{\overline{g}_0(\overline{\xi})}{\overline{\xi}}
    + \overline{g}_0'(\overline{\xi})
    \right] .
\end{equation}

According to our argument, the new backreaction criterion is realized when the slow-roll term (the unity) becomes comparable to the gauge-field correction, which implies
\begin{equation}
    b \simeq q\sqrt{6}
    \left[
    \frac{\overline{g}_0(\overline{\xi})}{\overline{\xi}}
    + \overline{g}_0'(\overline{\xi})
    \right]^{-1/2} ,
\end{equation}
where $q=\mathcal{O}(1)$ is a proportionality constant to be determined by fitting to the numerical results. To further simplify the expression, we use the large-$\overline{\xi}$ approximation for $\overline{g}_0(\overline{\xi})$ from Eq.~\eqref{eq:g0-approx}, which yields
\begin{equation}
\label{eq:b-of-xi-new}
    b \simeq q\sqrt{\frac{3}{\pi C}}\,
    \overline{\xi}^2 e^{-\pi \overline{\xi}}\, .
\end{equation}
Here, we have neglected corrections of order $\mathcal{O}(1/(2\pi\overline{\xi}))$, which are small for $\overline{\xi}\gtrsim 4$ in the regime of interest. 
Fitting Eq.~\eqref{eq:b-of-xi-new} to the exact numerical result in the region $\overline{\xi}>4$ gives $q\approx 1.43$, confirming that it is indeed of order unity. 
The total numerical prefactor in Eq.~\eqref{eq:b-of-xi-new} is therefore $q\sqrt{3/\pi C}\approx 86.8$. 

Inverting Eq.~\eqref{eq:b-of-xi-new}, we obtain the threshold value of $\overline{\xi}$ as a function of $b$:
\begin{equation}
   \label{eq:fit-xi-new-BR}
    \overline{\xi}_\mathrm{thr}^{\mathrm{new}}(b)
    \approx -\frac{2}{\pi}
    \mathcal{W}_{-1}\!\left(
    -\frac{\pi}{2}
    \sqrt{\frac{b}{86.8}}
    \right) .
\end{equation}
Finally, using the Klein--Gordon equation~\eqref{eq:stationary-solutions}, we obtain the corresponding expression for the threshold value of the gradient:
\begin{equation}
\label{eq:fit-v-new-BR}
    v_\mathrm{thr}^{\mathrm{new}}(b)
    = \frac{6\,\overline{\xi}_\mathrm{thr}^{\mathrm{new}}(b)}{b}+b\,\overline{g}_0\big[\overline{\xi}_\mathrm{thr}^{\mathrm{new}}(b)\big]\simeq \frac{6}{b}\left[\,\overline{\xi}_\mathrm{thr}^{\mathrm{new}}(b)+\frac{q^2}{2\pi}\right]
    \approx \frac{12}{\pi b}
    \left[-\mathcal{W}_{-1}\!\left(
    -\frac{\pi}{2}
    \sqrt{\frac{b}{86.8}}
    \right)+0.51 \right]\, ,
\end{equation}
where, in the second approximate equality, we used the asymptotic form of $\overline{g}_0(\overline{\xi})$ in Eq.~\eqref{eq:g0-approx} and also Eq.~\eqref{eq:b-of-xi-new}.
Equations~\eqref{eq:fit-xi-new-BR} and~\eqref{eq:fit-v-new-BR} provide an accurate fit to the red solid curves in Fig.~\ref{fig:thresholds}(a) and~\ref{fig:thresholds}(b), respectively, in the regime $b\ll 1$. In the figure, these approximations are shown by the black dashed-dotted lines.

The threshold value of the instability parameter, $\overline{\xi}_{\rm thr}$, depends only logarithmically on $b$ for both the old and the new criteria. 
Indeed, Eqs.~\eqref{eq:fit-xi-old-BR} and~\eqref{eq:fit-xi-new-BR} share the same leading behavior, $\overline{\xi}_{\rm thr} \simeq -\pi^{-1}\ln b + \text{const.}$
As a consequence, the corresponding threshold value of the potential-gradient parameter scales approximately as $v_{\mathrm{thr}}(b)\propto 1/b$.
More specifically, Eq.~\eqref{eq:fit-v-old-BR} gives 
$v_{\mathrm{thr}}^{\mathrm{old}}(b)=12\,\overline{\xi}_{\mathrm{thr}}^{\mathrm{old}}/b$, 
whereas Eq.~\eqref{eq:fit-v-new-BR} yields 
$v_{\mathrm{thr}}^{\mathrm{new}}(b)=6\,\overline{\xi}_{\mathrm{thr}}^{\mathrm{new}}/b$. 
Therefore, in the regime $b\ll 1$, the functions $v_{\mathrm{thr}}^{\mathrm{old}}(b)$ and $v_{\mathrm{thr}}^{\mathrm{new}}(b)$ differ approximately by a factor of two. 
This accounts for the vertical separation between the blue and red solid curves in Figs.~\ref{fig:backreaction} and~\ref{fig:thresholds}(b).
As $b$ increases, however, the two curves gradually approach each other and eventually coincide at the point indicated by the thick red dot in Figs.~\ref{fig:instability1}, \ref{fig:backreaction}, and~\ref{fig:thresholds}. 
This point corresponds to the codimension-2 bifurcation discussed in the next subsection.

\subsubsection*{New criterion: $\boldsymbol{\operatorname{Re}\zeta_1=0}$ at the SB--UB boundary}

We finally turn to the dashed red line in Figs.~\ref{fig:instability1} and~\ref{fig:backreaction}, which marks the separation between the SB and UB regimes. 
For large $b\gg 1$, this curve corresponds to an approximately constant value of $\overline{\xi}$. This behavior is evident in Fig.~\ref{fig:instability1}, where the red dashed line runs nearly parallel to the contours of constant $\overline{\xi}$. 
Although we do not currently have an analytical explanation for this feature, we adopt it as an empirical observation and proceed to construct suitable fit functions.

As discussed above, the threshold value of the instability parameter can be accurately approximated by a constant,%
\footnote{This value is close to $\pi$, but not exactly equal to it. 
A more accurate numerical estimate is $\xi_{\infty} \approx 3.1374$.}
\begin{equation}
\label{eq:fit-xi-stable-BR}
    \overline{\xi}_{\mathrm{thr}}^\mathrm{SB}
    = \xi_{\infty}
    \approx 3.14\, .
\end{equation}
Using the Klein--Gordon equation~\eqref{eq:stationary-solutions} and neglecting the first term (since along the red dashed curve $\delta_{\mathrm{KG}}\gg 1$, i.e., it lies well above the blue solid line), we obtain
\begin{equation}
\label{eq:fit-v-stable-BR}
    v_{\mathrm{thr}}^\mathrm{SB}
    = b\, \overline{g}_0(\xi_\infty)
    \approx 1220\, b\, .
\end{equation}
The fitting functions given in Eqs.~\eqref{eq:fit-xi-stable-BR} and~\eqref{eq:fit-v-stable-BR} are shown by the black dotted lines in panels~(a) and~(b) of Fig.~\ref{fig:thresholds}, respectively.


\subsection{Bifurcation diagram}
\label{subsec:bifurcation}

\begin{figure}[t!]
\centering
\includegraphics[width=0.98\linewidth]{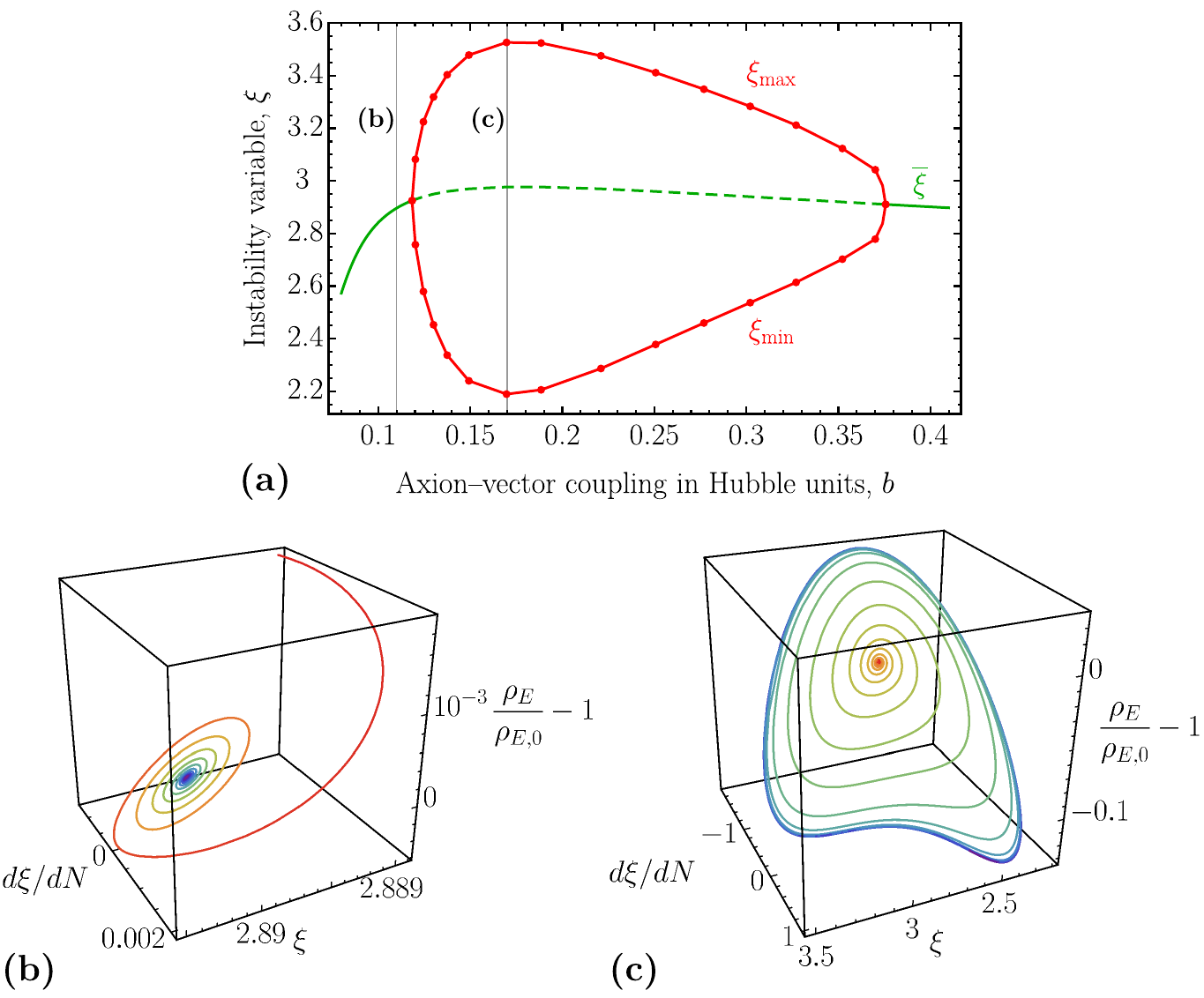}
\caption{Panel (a): Bifurcation diagram illustrating the transformation of a stable solution $\xi=\overline{\xi}(b,v)$ into an unstable one, followed by the emergence of a new stable solution, as the axion--vector coupling $b=\beta H/M_{\mathrm{P}}$ is gradually increased for a fixed gradient parameter $v=200$ (the parameters vary along the green segment shown in Figs.~\ref{fig:instability1} and \ref{fig:backreaction}). 
The solid green line shows the dependence $\overline{\xi}(b,v=200)$ when it corresponds to a stable solution, while the dashed green line indicates the same branch when it is unstable. The solid red curves show the maximum (upper) and minimum (lower) values of $\xi$ attained along the limit cycle within the unstable region. Red dots denote values extracted directly from numerical solutions of the full GEF system with truncation order $n_{\mathrm{tr}}=99$, while the red straight segments between them are obtained by interpolation. 
The figure demonstrates two supercritical Hopf bifurcations facing each other on opposite sides of the instability region. The two vertical solid lines mark $b=0.11$ and $b=0.17$, for which the phase-space trajectories are shown in panels (b) and (c), respectively.\\
Panels (b) and (c): Phase-space trajectories in the three-dimensional space spanned by $(\xi,\, d\xi/dN,\, \rho_E/\rho_{E,0}-1)$, starting from initial conditions slightly displaced from the stationary solution $\overline{\xi}(b,v)$.  As time increases, the trajectories are followed in a clockwise direction (color changing from red to purple). Panel (b) shows convergence toward a stable fixed point, whereas panel (c) shows convergence toward a limit cycle after the trajectory escapes the unstable fixed point.
\label{fig:bifurcation}}
\end{figure}

While the Lyapunov exponents identify the onset of instability, they do not describe the qualitative nature of the resulting dynamics. In this subsection, we study the evolution beyond the stability boundary. For this, we fix the gradient parameter at $v=200$ and vary the axion--vector coupling $\beta$, moving along the green segment in the stability diagram shown in Fig.~\ref{fig:instability1}. 
This yields the bifurcation diagram shown in Fig.~\ref{fig:bifurcation}(a). 
For each value of the coupling, the system is initialized slightly away from the stationary solution $\overline{\xi}(b,v)$ and evolved for a sufficiently long time so that transient effects decay and the trajectory reaches the asymptotic invariant manifold in phase space (either a fixed point or a limit cycle). The extrema of $\xi(N)$ on this manifold are then recorded, allowing us to characterize the long-term dynamical behavior. The approaches to the fixed point and to the limit cycle are illustrated in the phase-space plots in Figs.~\ref{fig:bifurcation}(b) and \ref{fig:bifurcation}(c), respectively. For visualization purposes, we employ a three-dimensional phase space spanned by $\xi$, its derivative $d\xi/dN$, and an additional independent dynamical variable (corresponding to $\delta e_0/\overline{e}_0$): the electric energy density $\rho_E = \langle \bm{E}^2 \rangle / 2$, normalized to its initial value in the simulation. This choice is not unique; any other variable linearly independent of the first two could be used equally well. Embedding the dynamics in a three-dimensional phase space is necessary to ensure that the system's trajectory does not self-intersect. This will be especially important for the more complex trajectories displayed in Fig.~\ref{fig:bursting} below.

At small coupling, all trajectories converge toward a unique stationary value $\bar{\xi}$, shown by the solid green curve in Fig.~\ref{fig:bifurcation}(a); see also Fig.~\ref{fig:bifurcation}(b) for the typical phase-space trajectory in this case. This corresponds to the stable constant-roll solution discussed earlier. 

As the coupling is increased, the stationary fixed point loses stability at a threshold value of the coupling when a complex conjugate pair of eigenvalues crosses the imaginary axis ($\mathrm{Re}\,\zeta_1=0$), i.e., at a \textit{Hopf bifurcation}.
Beyond this point, $\xi(N)$ becomes time-dependent and oscillates between a minimum $\xi_{\min}$ and a maximum $\xi_{\max}$. The width of the region between $\xi_{\min}$ and $\xi_{\max}$ grows continuously as the coupling increases, indicating a \textit{supercritical} Hopf bifurcation~\cite{Strogatz-Book:2015}. The phase-space trajectory corresponding to the maximal excursion in $\xi$ attainable for $v=200$ is shown in Fig.~\ref{fig:bifurcation}(c). The limit cycle, shown in purple, still has a relatively simple shape; cf.~the more intricate limit cycles that arise for larger values of $v=\mathcal{O}(10^7)$, discussed in Sec.~\ref{subsec:bursting}.

Upon increasing the coupling strength further, the oscillation amplitude decreases again until the system reenters the stable region. Qualitatively, the same behavior as before is observed, but in reverse order. The dynamical range of $\xi$ gradually decreases and, at the edge of the stability region, when $\mathrm{Re}\,\zeta_1=0$, it reduces to a point lying on the curve $\xi=\overline{\xi}(b,v=200)$ corresponding to the stationary solution. Thus, we observe two supercritical Hopf bifurcations facing each other in the diagram.

Note that as the value of $v$ decreases, the width of the instability region in Fig.~\ref{fig:instability1} gradually shrinks. At the critical value $v = v_c \approx 102$, the two bifurcations annihilate each other. This point corresponds to \textit{a codimension-2 Hopf bifurcation}~\cite{Kuznetsov:1998}, denoted by the red dot in Figs.~\ref{fig:instability1} and \ref{fig:backreaction}. Remarkably, this point lies exactly on the blue solid curve in Fig.~\ref{fig:backreaction}; that is, at this point (and only at this point) both the old and the new backreaction criteria are simultaneously satisfied. 
This statement follows from two observations. 
First, as discussed in Sec.~\ref{subsec:criterion}, the blue curve consists of the points where $\overline{\xi}$ as a function of $b$ along vertical slices at fixed $v$ attains its maximum in Fig.~\ref{fig:backreaction}, $\overline{\xi}_{\mathrm{max}}(v) = \max_b \overline{\xi}(v,b) = \overline{\xi}(v,b_{\rm max})$.
Second, our numerical LGEF analysis shows that the two bifurcation points always occur at $b$ values located on opposite sides of $b_{\mathrm{max}}$. 
As one approaches the codimension-2 bifurcation point, the two bifurcations move toward each other and eventually merge precisely at $b_{\mathrm{max}}$.

Sufficiently far from the codimension-2 bifurcation point, the Hopf bifurcations may in principle become subcritical. In that case, the red curves in Fig.~\ref{fig:bifurcation}(a), corresponding to the limit cycle, would extend into the range of the parameter $b$ where the fixed point remains stable (such that, for some values of $b$, the red lines would coexist with the solid green line), while the limit cycle itself would be unstable. At the bifurcation point, the stability would then flip: the fixed point (green curve) would lose its stability, and the limit cycle\,---\,already of finite amplitude\,---\,would become stable. However, in the present study, the Hopf bifurcations remain supercritical at least up to values of $v \sim 10^7$.


\subsection{Bursting dynamics}
\label{subsec:bursting}

Deeper within the unstable regime (for larger values of the gradient parameter $v$), the oscillatory solutions identified in the bifurcation diagram become strongly nonlinear.
Instead of smooth, nearly sinusoidal motion [see Fig.~\ref{fig:bifurcation}(c)], the system may even develop an intermittent bursting structure characterized by the coexistence of multiple time scales.

\begin{figure}
\centering
\includegraphics[width=0.97\linewidth]{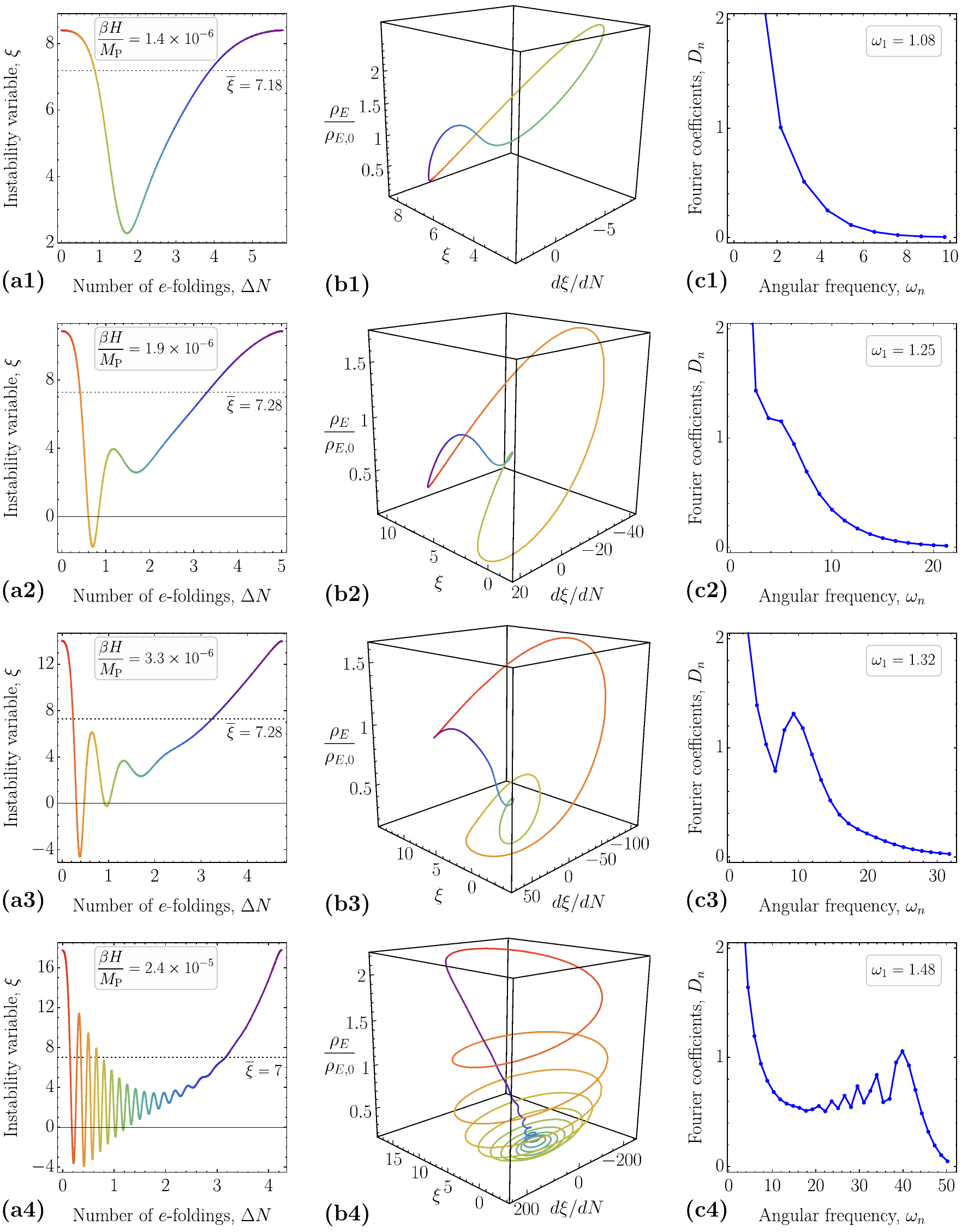}
\caption{Column (a): Time evolution of the instability variable $\xi$ during one period of motion on the late-time attractor.
Column (b): Projection of the late-time attractor (limit manifold) in the three-dimensional phase space spanned by $(\xi,\,d\xi/dN,\,\rho_E/\rho_{E,0})$; the color coding matches column (a). Column (c): Fourier spectra of the periodic solutions shown in column (a), with the fundamental frequency $\omega_1 = 2\pi/T$. Rows 1 to 4 correspond to the four parameter points marked by purple stars in Figs.~\ref{fig:instability1} and \ref{fig:backreaction}, for $v=3.6\times 10^7$ and $b \in \{1.4,\,1.9,\,3.3,\,24\}\times10^{-6}$, respectively.}
\label{fig:bursting}
\end{figure}

This transition is illustrated in Fig.~\ref{fig:bursting}. Column (a) shows the evolution of the instability variable $\xi(N)$, column (b) displays the corresponding phase-space trajectories, and column (c) presents the Fourier spectra.
The four rows correspond to increasing values of the axion--vector coupling. Specifically, the potential parameter is fixed to $v=3.6\times10^7$, while the coupling takes the values
$b = 1.4\times10^{-6},\, 1.9\times10^{-6},\,
3.3\times10^{-6}$, and $2.4\times10^{-5}$.
The corresponding parameter points are indicated by purple stars in Figs.~\ref{fig:instability1} and \ref{fig:backreaction}.

At late times, the system settles into a periodic attractor. One period of the evolution of $\xi$ is shown in column (a), where the color gradient indicates the progression of the time variable $N$. The same color coding is used in column (b) to trace the corresponding trajectory in the three-dimensional phase space spanned by $\xi$, its derivative $d\xi/dN$, and the electric energy density $\rho_E$, normalized to its initial value in the Anber--Sorbo solution.

The origin of this periodic behavior can be traced to the delayed response of gauge-field production to changes in the inflaton velocity, as previously discussed in Refs.~\cite{Notari:2016npn,Domcke:2020zez}.
Whenever the instability variable $\xi$ becomes sufficiently large, the tachyonic instability of the gauge field is strongly enhanced, leading to rapid particle production. The expectation value $\langle\bm{E}\cdot\bm{B}\rangle$ then grows rapidly and induces a substantial effective friction term in the axion equation of motion, Eq.~\eqref{KGF-2}. This abruptly suppresses $\dot{\phi}$ and terminates gauge-field production. As the produced gauge fields redshift away, the axion resumes its motion until the next cycle begins.

As the coupling increases further, the periodic evolution becomes
progressively more complex. Multiple fast oscillations develop within each period of the large-scale motion, and the structure of the attractor in phase space correspondingly changes. At smaller coupling, the limiting manifold is close to a weakly deformed ellipse. With increasing $b$, the trajectory winds several times before completing a single period. After a sequence of fast oscillations, the system undergoes a rapid, burst-like excursion and enters the next cycle. 
Importantly, this bursting regime is entirely deterministic and persists indefinitely. It arises from the nonlinear feedback between the axion velocity and the history-dependent amplification of gauge fields, rather than from stochastic effects or quantum fluctuations. At the same time, we remind the reader that our discussion pertains to our toy model in de Sitter space and neglects any spatial inhomogeneities in the axion field. Both deviations from pure de Sitter and axion gradients will modify the picture in Fig.~\ref{fig:bursting} in realistic scenarios. 

The emergence of an additional, shorter time scale is also evident in the Fourier spectra shown in column (c) of Fig.~\ref{fig:bursting}. To quantify this structure, we expand $\xi(N)$ in a discrete Fourier series. For a periodic solution with period $T$, we define the frequencies $\omega_n = 2\pi n/T$ ($n\in\mathbb{Z}_+$) and the corresponding Fourier coefficients,
\begin{align}
    A_n &= \frac{2}{T}\int_0^T dN\,\xi(N)\cos(\omega_n N)\,,\\
    B_n &= \frac{2}{T}\int_0^T dN\,\xi(N)\sin(\omega_n N)\,,\\
    D_n &= \sqrt{A_n^2 + B_n^2}\,. \vphantom{\int_0^T}
\end{align}
In all cases, the dominant peak occurs at the fundamental frequency $\omega_1$, corresponding to the large-scale periodic modulation. In all four panels, the value of $\omega_1$ is slightly larger than unity, which corresponds to a trajectory period of $T \simeq 5\text{--}6$ $e$-folds. This well-known result was first noticed in Ref.~\cite{Notari:2016npn}, explained in Ref.~\cite{Domcke:2020zez}, and subsequently investigated in Refs.~\cite{Peloso:2022ovc,vonEckardstein:2023gwk}, where these large-scale oscillations were attributed to the delayed response of the gauge field to variations in the inflaton velocity. 
Notably, the frequencies of the large-scale oscillations in the late-time nonlinear regime, $\omega_1$, are in good agreement with the imaginary parts of the corresponding Lyapunov exponents in the linear perturbation regime, $\mathrm{Im}\,\zeta_1$, which take the values $0.95$, $1.18$, $1.34$, and $1.49$ for panels (c1)--(c4) of Fig.~\ref{fig:bursting}, respectively. Thus, the linear perturbation analysis accurately predicts the oscillation frequency in the nonlinear regime.

As the coupling increases, an additional peak emerges and shifts toward higher frequencies, signaling the presence of the shorter time scale associated with the burst-like dynamics. It may get shorter and shorter with increasing coupling $b$ possibly even going to zero. At present, the physical origin of this short time scale and its behavior for larger $b$ remains unclear and deserves further investigation.

Note that, taking the decreasing characteristic short time scale as the unit of time, the evolution of the system from row~1 to row~4 in Fig.~\ref{fig:bursting} exhibits the following qualitative features: (i) the length of the phase-space trajectory rapidly increases; (ii) the period (in the rescaled time units) grows quickly. Both would diverge if the original characteristic short time approached zero. Such a behavior would correspond to the so-called \textit{blue-sky catastrophe}~\cite{Kuznetsov:1998,Shilnikov:2014bsc}, in which, at a finite value of a control parameter, both the period and the length of a limit cycle diverge. Beyond this critical point the periodic orbit would cease to exist.
However, for the considered value of the gradient parameter $v = 3.6\times 10^{7}$, both the trajectory length in phase space and the rescaled period remain finite and increase approximately linearly in the accessible range of the coupling $b$. 
Although we are not able to explore values of $b$ beyond the one in the fourth row of Fig.~\ref{fig:bursting} (since the computations become increasingly demanding), the occurrence of a blue-sky catastrophe appears unlikely for this value of $v$. Nevertheless, it cannot be ruled out for larger values of $v$, farther away from the codimension-2 bifurcation point.


\section{Application to realistic inflationary models} 
\label{sec:realistic}

In this section, we consider the full time-dependent inflationary background in two realistic inflationary models and demonstrate how the new criterion for strong backreaction, introduced in Sec.~\ref{subsec:Lyapunov}, can be applied in practice.

As a first example, we consider chaotic inflation with a quadratic potential,
\begin{equation}
\label{eq:potential-phi2}
    V_1(\phi) = \frac{m^2 \phi^2}{2}\, ,
\end{equation}
where, for definiteness, the inflaton mass is fixed to $m = 6\times10^{-6}\,M_{\mathrm{P}}$. Although this model is excluded by current CMB observations~\cite{Planck:2018jri} due to its prediction of a large tensor-to-scalar ratio, it remains useful as a benchmark. In particular, (i)~it accurately describes the shape of many inflationary potentials near the end of inflation, and (ii)~it has been widely used in previous studies of gauge-field production in axion inflation, both semi-analytical and lattice simulations.

As a second example, we consider the T-model of $\alpha$-attractors with the potential~\cite{Kallosh:2013hoa}
\begin{equation}
\label{eq:potential-tanh}
    V_2(\phi) =
    V_0 \tanh^2\!\left(\frac{\phi}{\sqrt{6\alpha} M_{\mathrm{P}}}\right),
\end{equation}
with $\alpha = 0.1$ and $V_0 = 10^{-11}\,M_{\mathrm{P}}^4$. It belongs to the class of plateau models, which are favored by the CMB observations~\cite{Martin:2024qnn}.

We now turn to a representation that enables us to follow the time evolution in realistic inflationary models. In principle, one could continue working in the $(v,b)$ parameter space. 
For practical applications, however, it is more convenient to consider the three-dimensional space spanned by the Hubble parameter $H$ and the slow-roll parameter $\epsilon_V = (M_\mathrm{P}^2/2)(V'/V)^2$, which characterize the inflationary background, together with the axion--vector coupling $\beta$ governing the interaction with the gauge field.
In the model-independent analysis of Sec.~\ref{sec:const_back}, it was sufficient to work with the two effective combinations $v$ and $b$, which fully determined the system’s dynamics. 
Within that framework, two models sharing the same values of $(v,b)$\,---\,even if corresponding to different triples $(H,\epsilon_V,\beta)$\,---\,were dynamically indistinguishable. 
In realistic scenarios, however, both $H$ and $\epsilon_V$ evolve during inflation according to the specific form of the inflaton potential. It is therefore natural to lift this degeneracy and enlarge the parameter space accordingly.

As discussed above, the system continuously moves through parameter space during its evolution. 
Nevertheless, over intervals of a few $e$-foldings the background spacetime is well approximated by quasi-de Sitter expansion. This justifies applying the stability diagram shown in Fig.~\ref{fig:instability1}, derived under the assumption of constant $H$ and $\epsilon_V$, locally along the inflationary trajectory. Our goal is to follow this trajectory until gauge-field backreaction becomes relevant, that is, until the condition $\operatorname{Re}\zeta_1 \ge 0$ is satisfied for the first time for a given value of $\beta$.

To determine this point, we use the expression for the threshold value of $v$ from Eq.~\eqref{eq:fit-v-new-BR}, together with the definition of $v$ in Eq.~\eqref{eq:def-tildev}. 
Solving the latter for $\epsilon_V$, we obtain contours in the $(H,\epsilon_V)$ plane corresponding to the onset of backreaction for fixed value of the axion--vector coupling $\beta$:
\begin{tcolorbox}
\vspace{-0.45cm}
\begin{equation}
\label{eq:epsilon-thr}
    \!\!\!\!\!\epsilon_V^{\mathrm{thr}}(H,\beta)
    = \frac{H^2}{18 M_\mathrm{P}^2}
    \left[
    v_{\mathrm{thr}}^\mathrm{new}
    \!\left(\frac{\beta H}{M_\mathrm{P}}\right)
    \right]^2 \approx \frac{8}{\pi^2\beta^2} 
    \left[\mathcal{W}_{-1}\!\left(
    -\frac{\pi}{2}
    \sqrt{\frac{\beta H/M_\mathrm{P}}{86.8}}
    \right)-0.51\right]^2 \,.
\end{equation}
\end{tcolorbox}

\begin{figure}
\centering
\includegraphics[height=6.6cm]{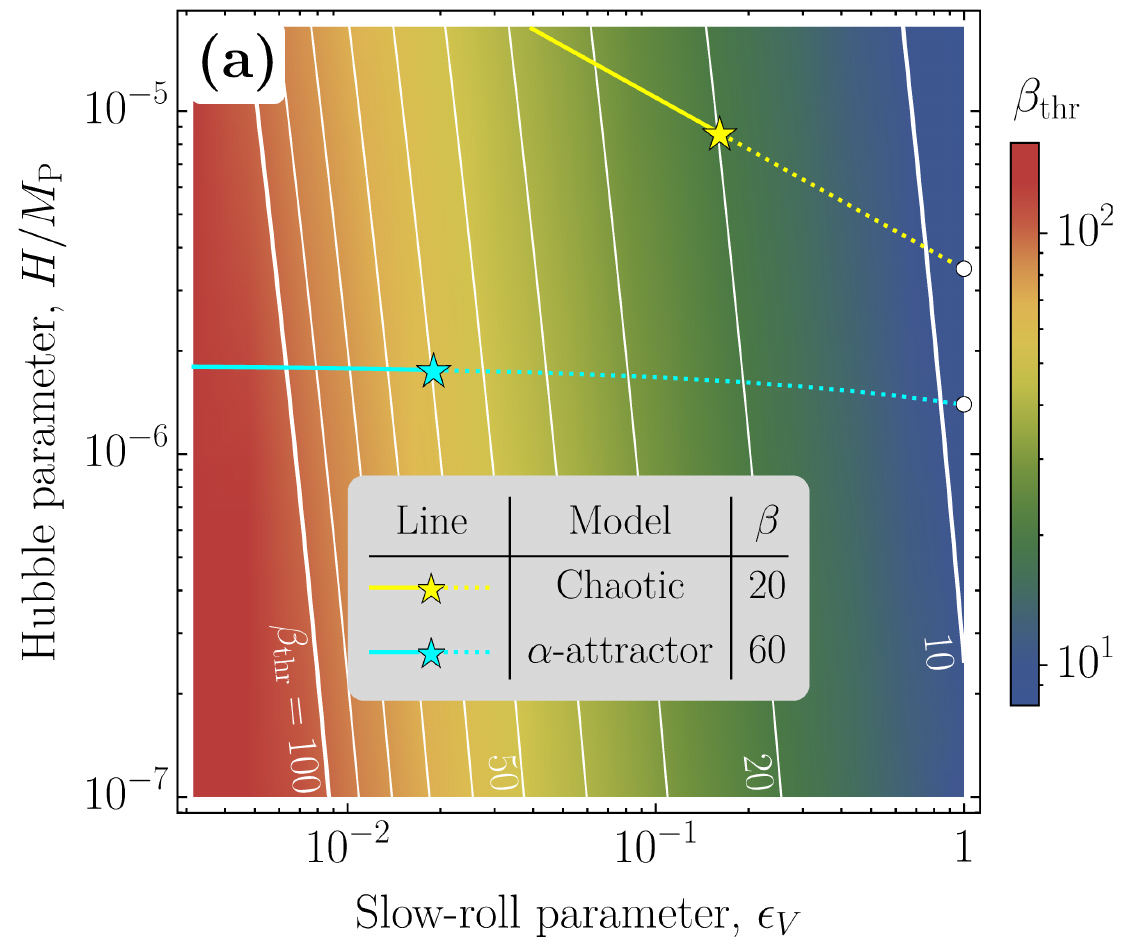}\hspace{3mm}
\includegraphics[height=6.6cm]{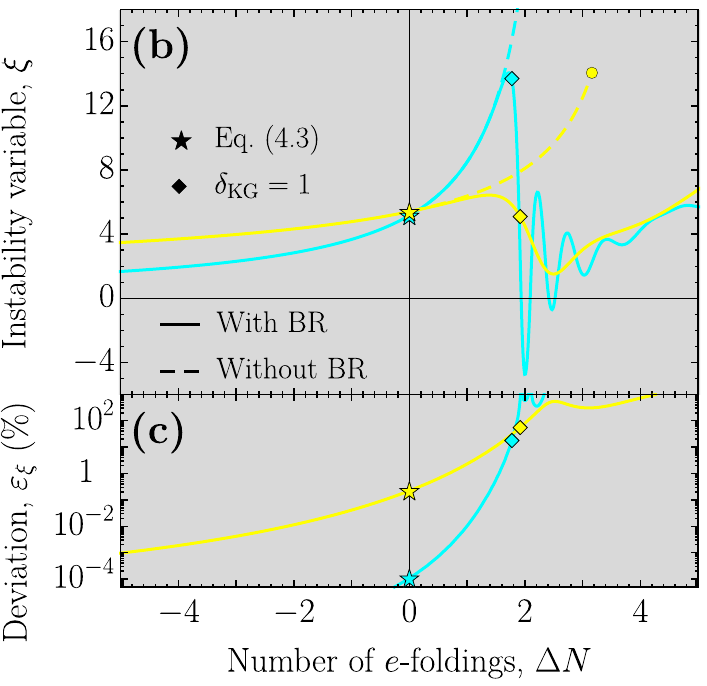}
\caption{Panel (a): Threshold value of the axion--vector coupling, $\beta_{\mathrm{thr}}$, at which the system enters the UB regime (corresponding to the solid red line in Fig.~\ref{fig:instability1}), shown as a function of the slow-roll parameter $\epsilon_V$ and the Hubble parameter $H$. The white curves denote contours of constant $\beta_{\mathrm{thr}}$. The yellow solid and dotted curves represent the slow-roll trajectory of chaotic inflation with the quadratic potential~\eqref{eq:potential-phi2} and $m=6\times10^{-6}\,M_{\mathrm{P}}$, while the cyan solid and dotted curves correspond to the $\alpha$-attractor $T$-model~\eqref{eq:potential-tanh} with $\alpha=0.1$ and $V_0=10^{-11}\,M_{\mathrm{P}}^4$. The yellow and cyan stars mark the intersections of the corresponding slow-roll trajectories with the contours $\beta_{\mathrm{thr}}=20$ and $\beta_{\mathrm{thr}}=60$, respectively, where the backreaction threshold in Eq.~\eqref{eq:epsilon-thr} is first satisfied for the respective models. In both cases, the trajectory is shown as a solid line prior to satisfying the backreaction criterion and continues as a dotted line afterwards.  Circle markers at the ends of the trajectories indicate the termination of the SR regime at $\epsilon_V=1$.
Panel (b): Instability variable $\xi$ as a function of the number of $e$-foldings $\Delta N$ for the chaotic inflation model with $\beta=20$ (yellow curves) and the $\alpha$-attractor T-model with $\beta=60$ (cyan curves). Solid curves show the solution of the full system including gauge-field backreaction, obtained using the GEF approach. Dashed curves correspond to the evolution with the gauge-field backreaction artificially switched off. For each model, the solid and dashed curves are horizontally shifted so that, at $\Delta N=0$, the inflaton value coincides with that of the slow-roll trajectory at the starred point in panel (a). Yellow and cyan diamonds mark the points along the solid curves where the condition $\delta_{\mathrm{KG}}=1$ is satisfied for the first time. The filled circle indicates the end of inflation on the dashed trajectory.
Panel (c): Relative deviation between the solid and dashed curves in panel (b), defined as $\varepsilon_\xi = \left[(\xi_{\mathrm{no\,BR}}/\xi_{\mathrm{GEF}})-1\right]\times 100\%$, plotted as a function of $\Delta N$. The stars, diamonds, and color coding follow the conventions of panel~(b).
\label{fig:models}}
\end{figure}

\noindent
Equivalently, for any given point $(H,\epsilon_V)$ one may determine the threshold coupling $\beta_{\rm thr}$ at which $\operatorname{Re}\zeta_1=0$ by solving Eq.~\eqref{eq:epsilon-thr} with respect to $\beta$. The color shading in Fig.~\ref{fig:models}(a) represents the resulting function $\beta_{\rm thr}(H,\epsilon_V)$, while the white contour lines indicate constant values of $\beta_{\rm thr}$ ranging from 10 to 100.
Note that the approximate expression given in the second equality of Eq.~\eqref{eq:epsilon-thr} reproduces the exact numerical result very accurately, with a relative error not exceeding 3.5\,\% over the entire parameter range shown in Fig.~\ref{fig:models}(a).

The expression for $\epsilon_V^{\mathrm{thr}}$ in Eq.~\eqref{eq:epsilon-thr} represents one our main results in this paper. It notably provides a convenient framework for analyzing realistic inflationary models. 
One simply overlays the slow-roll trajectory in the $(H,\epsilon_V)$ plane and identifies its intersection with the contour corresponding to the chosen value of $\beta$. 
By the slow-roll trajectory we mean the parametric curve $\epsilon_V(H)$ determined by
\begin{equation}
\label{eq:SR-trajectory}
    H(\phi) = \sqrt{\frac{V(\phi)}{3M_\mathrm{P}^2}}\, ,
    \qquad
    \epsilon_V(\phi) 
    = \frac{M_\mathrm{P}^2}{2}
    \left[
    \frac{V'(\phi)}{V(\phi)}
    \right]^2 .
\end{equation}
For a specified inflaton potential $V(\phi)$, the slow-roll trajectory can thus be obtained directly, without solving any differential equations.

As an illustration, Fig.~\ref{fig:models}(a) displays representative slow-roll trajectories for the two inflationary potentials introduced above. The chaotic inflation model is shown by the yellow curve, while the T-model trajectory is shown in cyan. The intersections of the slow-roll trajectories with the contours corresponding to the chosen values of $\beta$ are marked by stars of the respective colors. Segments of the trajectories prior to these intersection points are shown as solid lines, whereas the segments beyond them are dotted, emphasizing that the system departs from the slow-roll trajectory once backreaction becomes important.

To make this departure explicit, we examine the time evolution of the instability variable $\xi$ for each model. We consider the exact dynamics in two cases: 
(i)~the gauge-field backreaction is artificially switched off in the Friedmann and Klein--Gordon equations, and 
(ii)~the backreaction is consistently included using the GEF approach. 
Figure~\ref{fig:models}(b) shows the evolution of $\xi$ as dashed curves in case~(i) and as solid curves in case~(ii), with the color coding matching that of panel~(a).%
\footnote{Note that, even in case~(i), the system does not exactly follow the dotted slow-roll curve in Fig.~\ref{fig:models}(a), since the latter is obtained from Eq.~\eqref{eq:SR-trajectory} under the slow-roll approximation applied along the entire trajectory. The exact solution without gauge-field backreaction can be expanded in a perturbative series of Hubble flow parameters, while Eq.~\eqref{eq:SR-trajectory} is only valid at leading order in this perturbative series. Therefore, the dashed curves in panel~(b) always have a small deviation from the ideal slow-roll path. To avoid confusing this intrinsic deviation from the leading-order slow-roll result with the effect of gauge-field backreaction, we plot in Fig.~\ref{fig:models}(b) the exact trajectory without backreaction rather than the approximate slow-roll one.}
The number of $e$-foldings is normalized such that, at $\Delta N=0$, the inflaton value reached by the system in both cases coincides with its value on the slow-roll trajectory at the starred point in panel~(a). These reference points are therefore also indicated by stars in panel~(b).
In the full GEF solution, shown by the solid curves in Fig.~\ref{fig:models}(b), one can explicitly determine when the old backreaction condition $\delta_{\mathrm{KG}}=1$ is satisfied for the first time. 
For the two models under consideration, these points are indicated by diamonds of the corresponding colors along the solid curves.

Figure~\ref{fig:models}(c) presents the relative deviation between the dashed and solid curves in panel~(b) as a function of the number of $e$-foldings. The plot shows that the deviation induced by the gauge-field backreaction does not occur instantaneously when the new backreaction criterion is satisfied, but rather develops over a short interval of order $\mathcal{O}(1)$ $e$-folding. 
Remarkably, for some models, the relative deviation at the onset of backreaction can be extremely small (for instance, $\sim 10^{-4}$ in the T-model case). 
This demonstrates that the new criterion is capable of identifying the onset of backreaction dynamics even when the effect is initially still ``invisible to the naked eye.''
In contrast, the condition $\delta_{\mathrm{KG}}=1$ signals a stage where backreaction is already strong and relative deviations of $\mathcal{O}(1)$ are present, as indicated by the diamond markers.

These examples demonstrate that the new backreaction criterion based on the largest real part of the Lyapunov exponent, evaluated for fixed $H$ and $\epsilon_V$, performs well across qualitatively different classes of inflationary models, including both large-field and plateau models. The yellow and cyan stars in Fig.~\ref{fig:models} notably mark ``points of no return'', beyond which the system enters the UB regime. At the same time, we emphasize that our GEF results do not become \textit{immediately} unreliable (because of the absence of axion gradients) as soon as these points are crossed. Earlier comparisons with numerical lattice simulations have shown that the GEF results can often still be trusted for a small number of $e$-foldings, i.e., until shortly before the first local minimum in the evolution of $\xi$ is reached~\cite{Figueroa:2023oxc,Figueroa:2024rkr,Sharma:2024nfu} (see also Ref.~\cite{Domcke:2023tnn}). For our purposes, the full GEF results Fig.~\ref{fig:models}(b) thus represent a sufficient proxy of the exact lattice results. Our new criterion hence provides a practical and efficient diagnostic for identifying regions of parameter space free from unstable backreaction, without the need to solve the full set of background equations.


\section{Conclusions}
\label{sec:conclusion}

In this work, we performed a comprehensive investigation of the nonlinear dynamics of axion inflation coupled to an Abelian gauge field, with particular emphasis on the regime of strong gauge-field backreaction. Using the gradient-expansion formalism (GEF) and its linearized counterpart (LGEF), we carried out a systematic exploration of parameter space and clarified the precise conditions under which backreaction becomes dynamically important.

In our analysis, we revisited the long-standing question of the stability of the Anber--Sorbo solution, in which the axion potential gradient is balanced by Hubble friction and gauge-field-induced friction. While previous studies established that this solution is unstable in large parts of the strong-backreaction regime, we demonstrated that this conclusion is not universal.
By computing the full spectrum of Lyapunov exponents for homogeneous, time-dependent perturbations, we identified a qualitatively new parameter region in which the Anber--Sorbo solution remains dynamically stable despite strong backreaction. This regime, which we dubbed \emph{stable backreaction}, was not reported previously in the literature.

A central result of this work is the construction of an instability map based on the maximal real part of the Lyapunov exponents $\zeta_n$. This map reveals that the onset of instability generally precedes the point at which the conventional backreaction criterion, based on the magnitude of the $\langle\bm{E}\cdot\bm{B}\rangle$ term in the Klein--Gordon equation for the inflaton, is satisfied.
As a consequence, we proposed a new and more stringent criterion for the onset of (unstable) backreaction, formulated as the condition $\max\, (\operatorname{Re}\zeta_n) \geq 0$.

Beyond the linear stability analysis, we also investigated the nonlinear dynamics inside the unstable regime. By varying the axion--vector coupling at fixed potential shape, we identified a pair of supercritical Hopf bifurcations at the boundaries of the instability region. Between these bifurcations, the system evolves toward nontrivial limit manifolds in phase space. Deep inside the unstable regime, the dynamics becomes strongly nonlinear and exhibits deterministic bursting behavior, characterized by intermittent episodes of intense gauge-field production separated by comparatively quiescent phases. The presence of multiple intrinsic time scales in this regime has been confirmed by Fourier analysis.

We further applied our new backreaction criterion to realistic inflationary models with time-dependent background quantities. For large-field (chaotic) inflation and plateau-type $\alpha$-attractor models, we showed that, when the inflationary trajectories cross the instability boundary during their evolution, the deviations from the slow-roll trajectory occur on a short time scale, typically within $\sim 1$ $e$-folding after crossing the threshold. These results demonstrate that the Lyapunov-based backreaction criterion is robust across qualitatively different classes of inflationary potentials and can be used as a practical diagnostic without solving the full evolution equations\,---\,see Eq.~\eqref{eq:epsilon-thr} for a ready-to-use fit formula.

Our analysis is restricted to a homogeneous axion field and to perturbations depending only on time. While this limitation precludes definitive statements about the growth of spatial inhomogeneities, we argue that the stability properties identified here are likely relevant for superhorizon modes of axion perturbations. In particular, the existence of a stable-backreaction (SB) regime suggests that strong gauge-field production does not necessarily imply rapid growth of axion inhomogeneities or a breakdown of the homogeneous description. However, the answer to this important question can only be obtained by a detailed study of perturbations in the SB regime, which we postpone to future work.


\vskip.25cm
\section*{Acknowledgements}
We thank Alexandros Papageorgiou, Marco Peloso, Lorenzo Sorbo, and Ander Urio Garmendia for insightful discussions and valuable comments that helped to improve this work. K.\,S.\ is an affiliate member of the Kavli Institute for the Physics and Mathematics of the Universe (Kavli IPMU) at the University of Tokyo and as such supported by the World Premier International Research Center Initiative (WPI), MEXT, Japan (Kavli IPMU). At an early stage of this project, O.\,S. was supported by the Philipp Schwartz Fellowship of the University of M\"{u}nster. The work of O.\,S.\ has received funding through the SAFE\,---\,Supporting At-Risk Researchers with Fellowships in Europe project, which is funded by the European Union. Views and opinions expressed are, however, those of the authors only and do not necessarily reflect those of the European Union, the European Research Executive Agency (REA). Neither the European Union nor the granting authority can be held responsible for them.


\appendix

\section{Gradient-expansion formalism}
\label{app:GEF}

The background dynamics of homogeneous axion inflation can be analyzed using the gradient-expansion formalism (GEF). This approach reformulates the dynamical evolution of the axion--gauge-field system in terms of bilinear expectation values of the electric and magnetic field operators.
\begin{subequations}
 \begin{align}
    \mathcal{F}_E^{(n)}&\equiv \phantom{-}\frac{a^4}{k_{\mathrm{h}}^{n+4}}\langle \bm{E} \cdot \operatorname{rot}^n \bm{E}\rangle = \int\limits_{0}^{k_{\mathrm{h}}(t)}\frac{d k}{k} \frac{a^2 k^{n+3}}{2 \pi^2 k_{\mathrm{h}}^{n+4}}  \sum_{\lambda}\lambda^n |\dot{A}_\lambda(t,k)|^2\, ,\\
    \mathcal{F}_G^{(n)}&\equiv -\frac{a^4}{k_{\mathrm{h}}^{n+4}}\langle \bm{E} \cdot \operatorname{rot}^n \bm{B}\rangle = \int\limits_0^{k_{\mathrm{h}}(t)} \frac{d k}{k} \frac{a k^{n+4}}{2 \pi^2 k_{\mathrm{h}}^{n+4}}\sum_{\lambda}\lambda^{n+1} \operatorname{Re}[\dot{A}_\lambda(t,k)A_\lambda^*(t,k)]\, , \nonumber \\
    \mathcal{F}_B^{(n)}&\equiv \phantom{-}\frac{a^4}{k_{\mathrm{h}}^{n+4}}\langle \bm{B} \cdot \operatorname{rot}^n \bm{B}\rangle = \int\limits_{0}^{k_{\mathrm{h}}(t)} \frac{d  k}{k}  \frac{k^{n+5}}{2 \pi^{2}k_{\mathrm{h}}^{n+4}} \sum_{\lambda}\lambda^n |A_\lambda(t,k)|^2 \, ,
 \end{align}
\label{eq: GEF Bilinears}%
\end{subequations}
where $k_{\mathrm{h}}(t)$ is given by Eq.~\eqref{eq:cutoff-momentum}, and the mode functions $A_\lambda(t,k)$ are the solutions to the mode equation~\eqref{eq:mode-equation-physical-time}.
The evolution of the bilinear functions defined above is governed by an infinite chain of ordinary differential equations:
\begin{subequations}
    \begin{equation}
        \dot{\mathcal{F}}_E^{(n)} + (4+n)\frac{d \ln k_{\mathrm{h}}}{d t} \mathcal{F}_E^{(n)}  + 2\frac{k_{\mathrm{h}}}{a}\mathcal{F}_G^{(n+1)} - 4H\xi \mathcal{F}_G^{(n)}  = \frac{1}{4\pi^2}\frac{d \ln k_{\mathrm{h}}}{d t} \sum\limits_{\lambda=\pm 1} \lambda^n E_\lambda(\xi)\, ,\label{eq: GEF - En}
    \end{equation}
    \begin{equation}
        \dot{\mathcal{F}}_G^{(n)} + (4+n)\frac{d \ln k_{\mathrm{h}}}{d t} \mathcal{F}_G^{(n)} - \frac{k_{\mathrm{h}}}{a}\left(\mathcal{F}_E^{(n+1)} - \mathcal{F}_B^{(n+1)}\right) - 2H\xi \mathcal{F}_B^{(n)} = \frac{1}{4\pi^2}\frac{d \ln k_{\mathrm{h}}}{d t} \sum\limits_{\lambda=\pm 1} \lambda^{n+1} G_\lambda(\xi)\, , \label{eq: GEF - Gn}
    \end{equation}
    \begin{equation}
        \dot{\mathcal{F}}_B^{(n)} + (4+n)\frac{d \ln k_{\mathrm{h}}}{d t} \mathcal{F}_B^{(n)} - 2\frac{k_{\mathrm{h}}}{a}\mathcal{F}_G^{(n+1)}  =  \frac{1}{4\pi^2}\frac{d \ln k_{\mathrm{h}}}{d t} \sum\limits_{\lambda=\pm 1} \lambda^n B_\lambda(\xi)\, , \label{eq: GEF - Bn}
    \end{equation}
    \label{eq: GEF}%
\end{subequations}
where $\xi$ is the instability variable defined in Eq.~\eqref{xi}, $k_\mathrm{h}$ is the UV cutoff momentum given in Eq.~\eqref{eq:cutoff-momentum}, and the functions $E_\lambda$, $G_\lambda$, and $B_\lambda$ are given by
\begin{subequations}
\begin{align}
    \label{eq: boundary-En}
    E_\lambda(\xi)&=\frac{e^{\lambda \pi \xi}}{4 \xi^2} \left| (2i |\xi| - i \lambda \xi ) W_{-i \lambda \xi, \frac{1}{2}}(-4 i |\xi|) + W_{1-i \lambda \xi, \frac{1}{2}}(-4 i |\xi|) \right|^2 \, ,\\
    \label{eq: boundary-Gn}
    B_\lambda(\xi)&=\frac{e^{\lambda \pi \xi}}{2 |\xi|} \operatorname{Re}\left\{W_{1-i \lambda \xi, \frac{1}{2}}(-4 i |\xi|) W_{i \lambda \xi, \frac{1}{2}}(4 i |\xi|)\right\} \, ,\\
    \label{eq: boundary-Bn}
    B_\lambda(\xi)&=e^{\lambda \pi \xi} \left| W_{-i \lambda \xi, \frac{1}{2}}(-4 i |\xi|) \right|^2\, .
\end{align}
\label{eq: Source Terms}%
\end{subequations}
The infinite system of coupled differential equations~\eqref{eq: GEF} can be approximately closed at an order $n_{\mathrm{tr}} \sim \mathcal{O}(100)$ by using the truncation condition~\cite{Domcke:2023tnn}
\begin{equation}
    \dot{\mathcal{F}}_{X}^{(n_{\mathrm{tr}}+1)} \simeq \sum_{l=1}^{L} (-1)^{(l-1)} \left(
    \begin{array}{c}
      L \\
      l
    \end{array}
  \right) 
  \dot{\mathcal{F}}_{X}^{(n_{\mathrm{tr}} - 2l + 1)} \,, \quad X = E,\, G,\, B \,,
\end{equation}
where $L\geq 1$ is an integer. Previously~\cite{Gorbar:2021rlt,Gorbar:2021zlr,Durrer:2023rhc}, the value $L=1$ was used in numerical computations; however, larger values $L\sim 5-10$ may lead to a more stable numerical result. 

These equations must be solved alongside the Klein--Gordon equation, Eq.~\eqref{eq:Klein-Gordon-final}, and (in the realistic inflation case) the first Friedmann equation, Eq.~\eqref{eq:Friedmann-final}, to obtain the evolution of all background quantities.

The gradient-expansion formalism is implemented in a new Python package: the \texttt{GEF Factory (GEFF)}~\cite{vonEckardstein:2025jug}, available at \href{https://github.com/richard-von-eckardstein/GEFF}{github.com/richard-von-eckardstein/GEFF}. The numerical results in Sec.~\ref{sec:realistic} were obtained using this package.

\section{Comparison to previous studies}
\label{app:comp-previous}

The stability of the Anber--Sorbo solution has been studied previously in Refs.~\cite{Peloso:2022ovc,vonEckardstein:2023gwk}. While Ref.~\cite{Peloso:2022ovc} derived general analytical results under a set of simplifying assumptions, Ref.~\cite{vonEckardstein:2023gwk} performed a numerical parameter scan to determine the survival time of the Anber--Sorbo solution as a function of model parameters.

\begin{figure}[t!]
\centering
\includegraphics[width=0.99\linewidth]{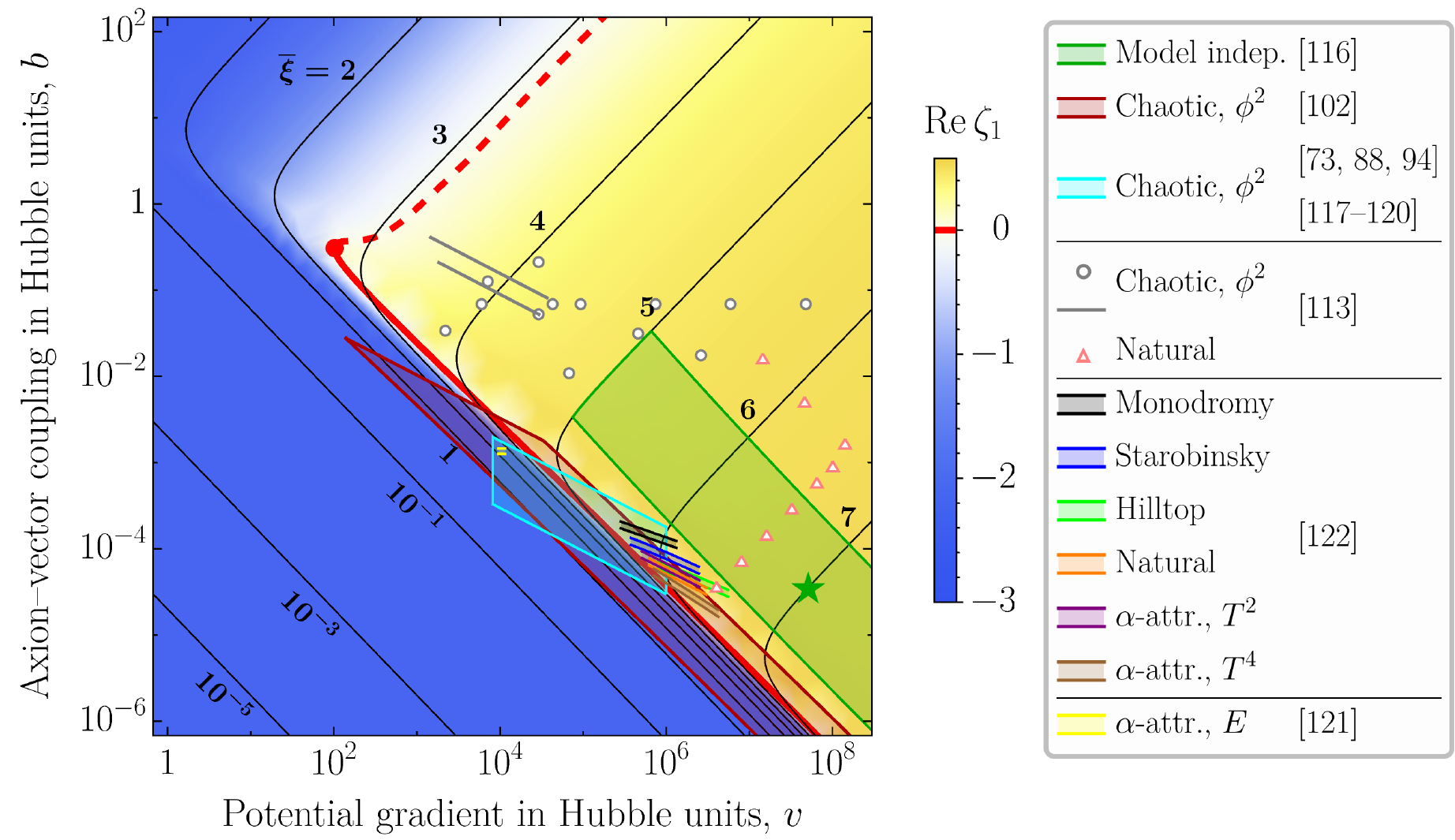}
\caption{The same stability diagram as in Fig.~\ref{fig:instability1}. Colored-shaded regions and markers indicate the parameter ranges and benchmark points studied in the works listed in the legend. In particular, the green-shaded band corresponds to the region of parameter space explored in Ref.~\cite{vonEckardstein:2023gwk} within a setup similar to the present study. The green star marks the central benchmark point, $v=3.6\times10^7$, $\beta=10^{2.5}$, and $\bar{\xi}=7$, for which most of the results of Ref.~\cite{vonEckardstein:2023gwk} were obtained. Models considered in other references are summarized in Table~\ref{tab:models-previous}.
\label{fig:instability2}}
\end{figure}

To reduce the dimensionality of parameter space, the authors of Ref.~\cite{vonEckardstein:2023gwk} fixed the Hubble rate during inflation to be a specific function of $\beta$ and $\overline{\xi}$,
\begin{equation}
\frac{H}{M_{\mathrm{P}}}
=
2 \times 10^{-7}
\left(\frac{100}{\beta}\right)^{1/2}
\exp\!\left[-2.85\,(\overline{\xi} - 7)\right] \, .
\end{equation}
This choice ensures that the backreaction term in the Klein--Gordon equation is already strong, $\delta_{\mathrm{KG}} \gg 1$, while the energy density of the gauge field remains negligible in the Friedmann equation.

For their numerical scan, the authors considered the parameter ranges $\beta \in [10^{1.5},\,10^{3.5}]$ and $\overline{\xi} \in [5,\,9]$. The corresponding region, translated into the $(v,\,b)$ parameter space used in the present work, is shown by the green-shaded rectangle in the stability diagram in Fig.~\ref{fig:instability2}.
The green star marks the benchmark point $\beta=10^{2.5}$, $\overline{\xi}=7$, which served as the primary reference point in Ref.~\cite{vonEckardstein:2023gwk}.

Several studies~\cite{Cheng:2015oqa,Domcke:2020zez,Gorbar:2021rlt,vonEckardstein:2025oic,Notari:2016npn,Caravano:2022epk,Figueroa:2023oxc,Sharma:2024nfu,Figueroa:2024rkr,Jamieson:2025ngu,Lizarraga:2025aiw} 
have investigated gauge-field production in realistic realizations of axion inflation using a variety of methods, including the mode-by-mode treatment in momentum space~\cite{Cheng:2015oqa,Domcke:2020zez,Notari:2016npn}, the GEF approach~\cite{Gorbar:2021rlt,vonEckardstein:2025oic}, and lattice simulations~\cite{Caravano:2022epk,Figueroa:2023oxc,Sharma:2024nfu,Figueroa:2024rkr,Jamieson:2025ngu,Lizarraga:2025aiw}. 
The inflaton potentials and parameter ranges explored in these works are summarized in Table~\ref{tab:models-previous}, and the corresponding regions in the $(v,b)$ parameter space are shown in Fig.~\ref{fig:instability2}.

\begin{table}
\centering\small
\begin{tabular}{c|c|c|c|c}
\hline\hline
Reference & Model & Potential $V(\phi)$ & Parameters & Method \\
\hline\hline
\multirow{2}{*}{\cite{Domcke:2020zez}}  & \multirow{21}{*}{Chaotic} 
      & \multirow{21}{*}{$m^2 \phi^2 / 2$} 
      & $m=6\times 10^{-6}M_{\mathrm{P}}$ & \multirow{2}{*}{Iterative MbM}\\ 
& & & $\beta \in \{20, 25\}$ & \\
      \cline{1-1} \cline{4-5}
\multirow{2}{*}{\cite{Cheng:2015oqa}} &  &  & $m=7.39\times 10^{-6}M_{\mathrm{P}}$ & \multirow{2}{*}{MbM} \\ 
& & & $\beta \in [7,32]$ & \\
\cline{1-1} \cline{4-5}
\multirow{2}{*}{\cite{Gorbar:2021rlt}}  &  &  &  $m=6\times 10^{-6}M_{\mathrm{P}}$ & \multirow{4}{*}{GEF} \\ 
& & & $\beta \in \{10, 20, 25\}$ & \\ \cline{1-1} \cline{4-4}
\multirow{2}{*}{\cite{vonEckardstein:2025oic}}  &  &  &  $m\in[10^{-20},10^{-5}]M_{\mathrm{P}}$ & \\
& & & $\beta \in [10, 55]$ & \\ \cline{1-1} \cline{4-5}
\multirow{2}{*}{\cite{Caravano:2022epk}} &  &  & $m=5.1\times 10^{-6}M_{\mathrm{P}}$  & \multirow{10}{*}{Lattice} \\ 
& & & $\beta \in \{25, 42\}$ & \\
\cline{1-1} \cline{4-4}
\multirow{2}{*}{\cite{Figueroa:2023oxc}} &  &  & $m=6.16\times 10^{-6}M_{\mathrm{P}}$  &  \\ 
& & & $\beta \in \{15, 18, 20\}$ & \\
\cline{1-1} \cline{4-4}
\multirow{2}{*}{\cite{Sharma:2024nfu}} &  &  & $m=5.3\times 10^{-6}M_{\mathrm{P}}$  & \\
& & & $\beta \in [6, 18]$ & \\ \cline{1-1} \cline{4-4}
\multirow{2}{*}{\cite{Figueroa:2024rkr}} &  &  & $m=6.16\times 10^{-6}M_{\mathrm{P}}$  & \\ 
& & & $\beta \in [10, 35]$ & \\ \cline{1-1} \cline{4-4}
\multirow{2}{*}{\cite{Lizarraga:2025aiw}} &  &  & $m=6.16\times 10^{-6}M_{\mathrm{P}}$  &  \\
& & & $\beta \in [14.25, 17.1]$ & \\ \cline{1-1} \cline{4-5}
\multirow{6}{*}{\cite{Notari:2016npn}} &  &  & $m\in[6.6,500]\times 10^{-6}M_{\mathrm{P}}$  & \multirow{6}{*}{MbM} \\ 
& & & $\beta \in [64, 1024]$ & \\
& & & $m=10^{-2}M_{\mathrm{P}}$, $\beta = 10^{-2}$ & \\
\cline{2-4}
& \multirow{6}{*}{Natural}& \multirow{6}{*}{$V_0\Big[1-\cos\frac{\phi}{v}\Big]$}& $V_0=[10^{-13},10^{-8}]M_\mathrm{P}^4$ & \\ 
& & & $v=[0.125,5]M_\mathrm{P}$ & \\
& & & $\beta = 167 \times (M_\mathrm{P}/v)$ & \\ \cline{1-1} \cline{4-5}
\multirow{18}{*}{\cite{Lizarraga:2025aiw}}& & &$V_0=5.9\times 10^{-10}M_\mathrm{P}^4$ & \multirow{21}{*}{Lattice} \\
& & & $v=\sqrt{8\pi}M_\mathrm{P}$ & \\
& & & $\beta\in[14.9,17.9]$ & \\ \cline{2-4}
& \multirow{3}{*}{Monodromy} & \multirow{3}{*}{$\mu^3(\sqrt{\phi^2+\phi_c^2}-\phi_c)$} & $\mu = 6\times 10^{-4}M_\mathrm{P}$ & \\
& & & $\phi_c=0.1M_\mathrm{P}$ & \\
& & & $\beta\in[18.9,22.7]$ & \\ \cline{2-4}
& \multirow{3}{*}{Hilltop} & \multirow{3}{*}{$V_0\Big[1-\Big(\frac{\phi}{v}\Big)^{\! 4\,}\Big]^2$} & $V_0 = 5.7\times 10^{-11}M_\mathrm{P}^4$ & \\
& & & $\phi_c=8M_\mathrm{P}$ & \\
& & & $\beta\in[20.8,25]$ & \\ \cline{2-4}
& \multirow{12}{*}{$\alpha$-attractors} & \multirow{3}{*}{$\frac{\Lambda^4}{2}\tanh^2\!\frac{\phi}{M}$} & $\Lambda = 7.16\times 10^{-3}M_\mathrm{P}$ & \\
& & & $M=8.79M_\mathrm{P}$ & \\
& & & $\beta\in[15.2,18.2]$ & \\ \cline{3-4}
& & \multirow{3}{*}{$\frac{\Lambda^4}{4}\tanh^4\!\frac{\phi}{M}$} & $\Lambda = 8.47\times 10^{-3}M_\mathrm{P}$ & \\
& & & $M=8.22M_\mathrm{P}$ & \\
& & & $\beta\in[12.6,15.1]$ & \\ \cline{3-4}
& & \multirow{6}{*}{$V_0\Big(1-e^{-\alpha_v \phi/M_{\mathrm{P}}}\Big)^2$} & $V_0 = 6.2\times 10^{-10}M_\mathrm{P}$ & \\
& & & $\alpha_v=3/10$ & \\
& & & $\beta\in[17.8,21.4]$ & \\ \cline{1-1} \cline{4-4}
\multirow{3}{*}{\cite{Jamieson:2025ngu}}& & & $V_0 = 1.08\times 10^{-11}M_\mathrm{P}$ & \\
& & & $\alpha_v=\sqrt{20/3}$ & \\
& & & $\beta\in[665,770]$ & \\
\hline\hline
\end{tabular}
\caption{Summary of axion-inflation models with gauge-field production considered in the literature, including the associated potentials, parameter ranges, and numerical methods employed. The corresponding regions in the $(v,b)$ parameter space are displayed in Fig.~\ref{fig:instability2}. \label{tab:models-previous}}
\end{table}

Importantly, all parameter regions considered in previous studies fall within the conventional slow-roll or unstable-backreaction regimes. In this work, we identify a new regime\,---\,the stable backreaction\,---\,that has not been explored before and exhibits qualitatively different behavior; see Sec.~\ref{subsec:Lyapunov}.
Within the assumptions of  Ref.~\cite{vonEckardstein:2023gwk}, the conclusion that the Anber--Sorbo solution is generically unstable was therefore correct. Our results demonstrate, however, that this conclusion does not hold universally once a broader region of parameter space is explored.




\end{document}